\documentclass[prl,onecolumn,amsmath,amssymb,superscriptaddress]{revtex4-2}
\usepackage{bm,psfrag,graphicx,setspace}
\usepackage{braket}
\usepackage[left]{lineno}
\usepackage{xcolor}
\selectfont

\DeclareFontFamily{OMS}{oasy}{\skewchar\font48 }
\DeclareFontShape{OMS}{oasy}{m}{n}{%
         <-5.5> oasy5     <5.5-6.5> oasy6
      <6.5-7.5> oasy7     <7.5-8.5> oasy8
      <8.5-9.5> oasy9     <9.5->  oasy10
      }{}
\DeclareFontShape{OMS}{oasy}{b}{n}{%
       <-6> oabsy5
      <6-8> oabsy7
      <8->  oabsy10
      }{}
\DeclareSymbolFont{oasy}{OMS}{oasy}{m}{n}
\SetSymbolFont{oasy}{bold}{OMS}{oasy}{b}{n}

\DeclareMathSymbol{\smallleftarrow}     {\mathrel}{oasy}{"20}
\DeclareMathSymbol{\smallrightarrow}    {\mathrel}{oasy}{"21}
\DeclareMathSymbol{\smallleftrightarrow}{\mathrel}{oasy}{"24}

\let\vec\mathbf

\begin{document}

\author{Marc R. Bourgeois}
\affiliation
{Department of Chemistry, University of Washington, Seattle WA, 98195}
\author{Austin G. Nixon}
\affiliation
{Department of Chemistry, University of Washington, Seattle WA, 98195}
\author{Matthieu Chalifour}
\affiliation
{Department of Physics, University of Washington, Seattle WA, 98195}
\author{David J. Masiello}
\email{masiello@uw.edu}
\affiliation
{Department of Chemistry, University of Washington, Seattle WA, 98195}

\title{Optical Polarization Analogs in Inelastic Free Electron Scattering}




\begin{abstract}
Advances in the ability to manipulate free electron phase profiles within the electron microscope have spurred development of quantum-mechanical descriptions of electron energy loss (EEL) processes involving transitions between phase-shaped transverse states. Here, we elucidate an underlying connection between two ostensibly distinct optical polarization analogs identified in EEL experiments as manifestations of the same conserved scattering flux. Our work introduces a procedure for probing  general tensorial target characteristics including global mode symmetries and local polarization.
\end{abstract}

\maketitle


Inherently necessitating the exchange of both energy and momentum, measurements involving the absorption and scattering of optical waves and energetic particles provide a wealth of information characterizing atomic, molecular, and nanoscale systems. Selection rules based on the optical polarization degrees of freedom, in particular, are indispensable tools for probing target excitation symmetries, albeit with spatial resolution fundamentally limited by diffraction. With their atomic-scale de Broglie wavelength, swift electrons in the scanning transmission electron microscope (STEM) offer superior spatial localization but lack polarization degrees of freedom since they are accurately described by the scalar Schr{\"o}dinger equation. Despite this deficiency, it was shown early on that transverse linear momentum transfer $\hbar{\bf q}_\perp$ in near-edge electron energy loss (EEL) processes could be exploited to selectively probe atomic inner shell excitations with distinct symmetries \cite{PhysRevLett.70.1822} and more recently to probe transversely polarized electric fields on the atomic length scale \cite{muller2014atomic}. A formal connection between the photon polarization $\hat{\bm\epsilon}$ in X-ray absorption measurements and $\hbar \vec{q}_{\perp}$ during core-loss inelastic electron scattering measurements was established in the electrostatic limit \cite{Kohl1985-fv, Hitchcock1993-vs, Yuan1997-hj, Hebert2003-mn, bradley2010comparative}, culminating in the experimental realization of magnetic circular dichroism measurements within a TEM \cite{Schattschneider2006-fy, rusz2014achieving}.

Improved monochromation and aberration correction technologies, on the other hand, have enabled STEM-EEL characterization of plasmonic \cite{nelayah2007mapping, smith2019direct}, nanophotonic \cite{polman2019electron, auad2021unveiling}, and phononic \cite{idrobo2018temperature, lagos2018thermometry, lagos2017mapping} systems in the low-loss ($\lesssim 10$ eV) regime with nanometer-scale spatial resolution. At such low energies \cite{de2010optical, garcia2021optical}, the STEM-EEL observable primarily probes the component of the generalized electromagnetic local density of optical states (EMLDOS) of the target specimen projected along the TEM axis \cite{de2008probing, hohenester2009electron}. However, following the demonstration of vortex electron beams carrying quantized orbital angular momentum (OAM) \cite{uchida2010generation, verbeeck2010production}, there has been considerable interest in studying OAM transfer between vortex free electron states and atomic \cite{lloyd2012quantized, van2015inelastic} as well as nanophotonic \cite{asenjo2014dichroism, cai2018efficient, zanfrognini2019orbital} targets. In particular, it was demonstrated that the symmetries of excited plasmonic modes could be controlled by pre- and post-selection of the transverse wave functions of the probing free electrons \cite{ugarte2016controlling, guzzinati2017probing}. More recently, a quasistatic theory was presented in which the transition dipole $\hat{\bf d}^{\perp}_{fi}$ arising during transitions between spatially localized phase-shaped transverse electron states plays the role of an optical polarization analog (OPA) in EEL processes, allowing access to additional components of the target's generalized EMLDOS \cite{lourencco2021optical}. Despite the coexistence of the $\hat{\bf d}^{\perp}_{fi}$ OPA in low-loss STEM-EEL measurements, and the $\hat{\vec{q}}_{\perp}$ OPA in core-loss scattering processes, the connection between these two OPAs has yet to be made explicit.

Here we present a general theoretical framework for describing fully retarded inelastic electron scattering and elucidate the notion of and relationships between OPAs in such measurements. Employing a formalism that explicitly accounts for the swift electron transverse degrees of freedom, we uncover an underlying connection between the two ostensibly distinct OPAs previously identified in linear momentum- (LM-) and OAM-resolved measurements under wide-field and focused beam conditions. Despite their apparent differences and regimes of applicability, the $\hat{\mathbf{q}}_{\perp}$ and $\hat{\bf d}^{\perp}_{fi}$ OPAs arising during LM and OAM transfer processes are both manifestations of the transverse components of the transition current density arising in our current-current response formalism. Numerical calculations highlighting the utility of phase-shaped EEL nanospectroscopy for determining mode symmetries and probing the 3D polarization-resolved response field of a plasmonic dimer target with nanoscale spatial resolution are presented.

 \begin{figure}
     \centering
     \includegraphics{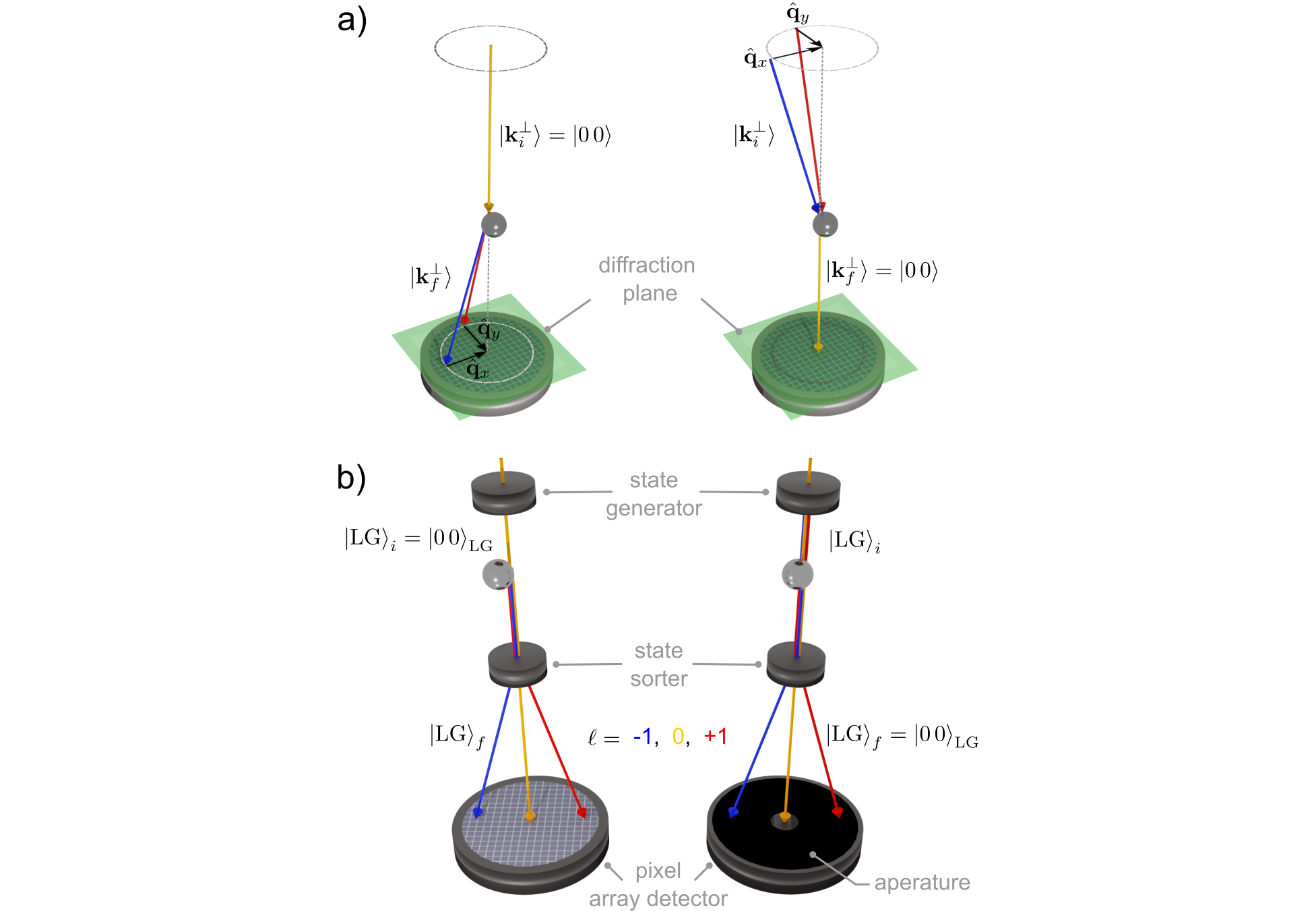}
     \caption{LM- and OAM-resolved EEL processes. Schemes showing (a) plane wave $\ket{0\,0} \rightarrow \ket{\vec{k}_f^{\perp}}$ (left) and $\ket{\vec{k}_i^{\perp}} \rightarrow \ket{0\,0}$ (right) LM-resolved and (b) Laguerre-Gauss $\ket{0\,0}_{\textrm{LG}} \rightarrow \ket{\ell\,p}_{\textrm{LG}}$ (left) and $\ket{\ell\,p}_{\textrm{LG}} \rightarrow \ket{0\,0}_{\textrm{LG}}$ (right) OAM-resolved EEL measurements.}
     \label{F1}
 \end{figure}

 Central to the present work are pre- and post-selected phase measurements, whereby the EEL signal is acquired for specific scattering channels determined by the transverse phase profiles of the initial and final free electron states. LM- and OAM-resolved EEL processes, shown schematically in Figure \ref{F1}, constitute two well-known examples of such a measurement. In its simplest realization, the LM-resolved measurement involves the preparation of an initial plane wave state $\ket{\vec{k}_i}$, which evolves into a different superposition of plane wave states via interaction with the target specimen. Post-selection of the final transverse state $\ket{\vec{k}_f}$ fixes the LM transfer $\hbar \vec{q} = \hbar (\vec{k}_i - \vec{k}_f)$ and can be accomplished via spatial filtering within the diffraction plane \cite{muller2014atomic, krehl2018spectral}. Figure \ref{F1}b depicts OAM-resolved EEL processes, where OAM state generators \cite{verbeeck2018demonstration, PhysRevResearch.2.043227, Madan2022-gj, Tsesses2023-hw} and sorters \cite{grillo2017measuring, tavabi2021experimental} perform selection of initial and final vortex states with well-defined OAM.

Within the first Born approximation, the rate of scattering from the initial light-matter state $ \ket{i} \ket{0}$ to a given final state $ \ket{f} \ket{n}$ is equal to that found using Fermi's golden rule with the interaction potential $\hat{V} = ({e}/{2mc})(\hat{\vec{A}} \cdot \hat{\vec{p}} + \hat{\vec{p}} \cdot \hat{\vec{A}})$ in the generalized Coulomb gauge \cite{glauber1991quantum} defined by $\nabla \cdot \varepsilon(\vec{x}) \vec{A}(\vec{x}, t) = 0$ with zero scalar potential. Using a mode expansion of the target's vector potential $\vec{A}(\vec{x}, t) = \sum_{n} \hat{a}^{\dagger}_n \vec{A}_{n}^{(-)}(\vec{x})e^{+i\omega_n t} + \hat{a}_n\vec{A}_{n}^{(+)}(\vec{x})e^{-i\omega_n t}$, the state-to-state frequency-resolved EEL transition rate becomes \cite{SM, lourencco2021bridging}
\begin{equation}
    \begin{aligned}
        w^{\textrm{loss}}_{fi}(\omega) 
        &= \frac{2 \pi}{ \hbar^2} \sum_{n} \bigg| \frac{1}{c} \int d\mathbf{x}'~\vec{A}^{(-)}_{n}(\vec{x}') \cdot \vec{J}_{fi}(\vec{x}')\bigg|^2 \delta(\omega - \omega_{if}) \delta(\omega - \omega_n) \\
        &= \frac{4 \pi^2}{ \hbar \omega} \int d  \vec{x} \, d\vec{x}'~ \vec{J}^{*}_{fi}(\vec{x}) \cdot \tensor{\boldsymbol\varrho}(\vec{x}, \vec{x}', \omega) \cdot \vec{J}_{fi}(\vec{x}') \delta(\omega - \omega_{if}),\\
    \end{aligned}
    \label{w_fi_loss}
\end{equation}
where $E_{if} = \hbar \omega_{if}$ is the energy difference between initial $\ket{i}$ and final $\ket{f}$ electron states, $\tensor{\boldsymbol{\varrho}}(\vec{x}, \vec{x}', \omega) = ({- 2 \omega}/{\pi}) \textrm{Im}\big\{ \tensor{\bf G}(\vec{x}, \vec{x}', \omega) \big\}$ is the EMLDOS tensor, and $\tensor{\bf G}(\vec{x}, \vec{x}', \omega)$ is the induced vector Helmholtz Green's tensor. The free electron transition current density $\vec{J}_{fi}(\vec{x}) = ({i \hbar e}/{2m})\{ \psi^*_f(\vec{x}) \nabla \psi_i(\vec{x})  - \psi_i(\vec{x}) \nabla \psi^*_f(\vec{x})\}$ is fully determined by the initial and final electron states and its orientation determines which components of the EMLDOS tensor contribute to $w_{fi}^{\textrm{loss}}(\omega)$ at each point in space. Eq. \eqref{w_fi_loss} is reminiscent of the optical plane wave extinction cross section $\sigma_{\textrm{ext}}(\omega) = 4 \pi({\omega}/{c}) \textrm{Im} \big\{  \hat{\boldsymbol{\epsilon}}^{*} \cdot \tensor{\boldsymbol{\alpha}}( \omega) \cdot \hat{\boldsymbol{\epsilon}} \big\}$ \cite{SM}, where $\tensor{\bm\alpha}(\omega)$ is the polarizability tensor characterizing the target response to plane wave excitation with polarization unit vector $\hat{\boldsymbol{\epsilon}}$. An arbitrary free photon pure polarization state is described by a point on the Poincar{\'e} sphere \cite{SM} with antipodal points $\{\hat{\boldsymbol{\epsilon}}_{1},\hat{\boldsymbol{\epsilon}}_{2},\hat{\boldsymbol{\epsilon}}_{+},\hat{\boldsymbol{\epsilon}}_{-}\}$ spanning the two-dimensional photon polarization Hilbert space.

If the free electron wave function can be separated within an orthogonal coordinate system with ${\bf x}=(x_1, x_2, x_3)$ and translational invariance along $x_3$, then the wave function can be written as $\psi(\vec{x}) = \Psi({\bf x}_\perp)e^{i k_3 x_3} = \Psi_1(x_1)\Psi_2(x_2)e^{i k_3 x_3}$ and the transition current density is $\vec{J}_{fi}(\vec{x})={\bf J}^{\perp}_{fi}({\bf x}_{\perp})e^{iq_{\parallel}x_3} +J^{\parallel}_{fi}(\vec{x}) \hat{\vec{x}}_3$, with $q_{\parallel} = k_3^i - k_3^f$. These conditions on $\psi(\vec{x})$ are satisfied within the Cartesian as well as polar, elliptical, and parabolic cylindrical coordinate systems \cite{gutierrez2000alternative}. We note \cite{SM} that $\mathbf{J}_{fi}(\vec{x})\cdot \hat{\bf x}_j$ ($j=1,2$) vanishes provided (1) $\Psi_j(x_j)$ remains unchanged during the interaction, and (2) $\textrm{Arg}\big\{ \Psi_j(x_j) \big\}$ is constant. Consequently, given the ability to perform EEL measurements with pre- and post-selection of free electron transverse states with sculpted phase profiles, OPAs can be defined by identifying pairs of initial and final states such that $\hat{\bf J}^{\perp}_{fi}({\bf x}_\perp)\to\hat{\bf J}^{\perp}_{fi}$ is position-independent and described by a point on the Poincar{\'e} sphere with antipodal points constructed from unit vectors $\hat{\bf x}_1$ and $\hat{\bf x}_2$.

\begin{figure}
    \centering
    \includegraphics{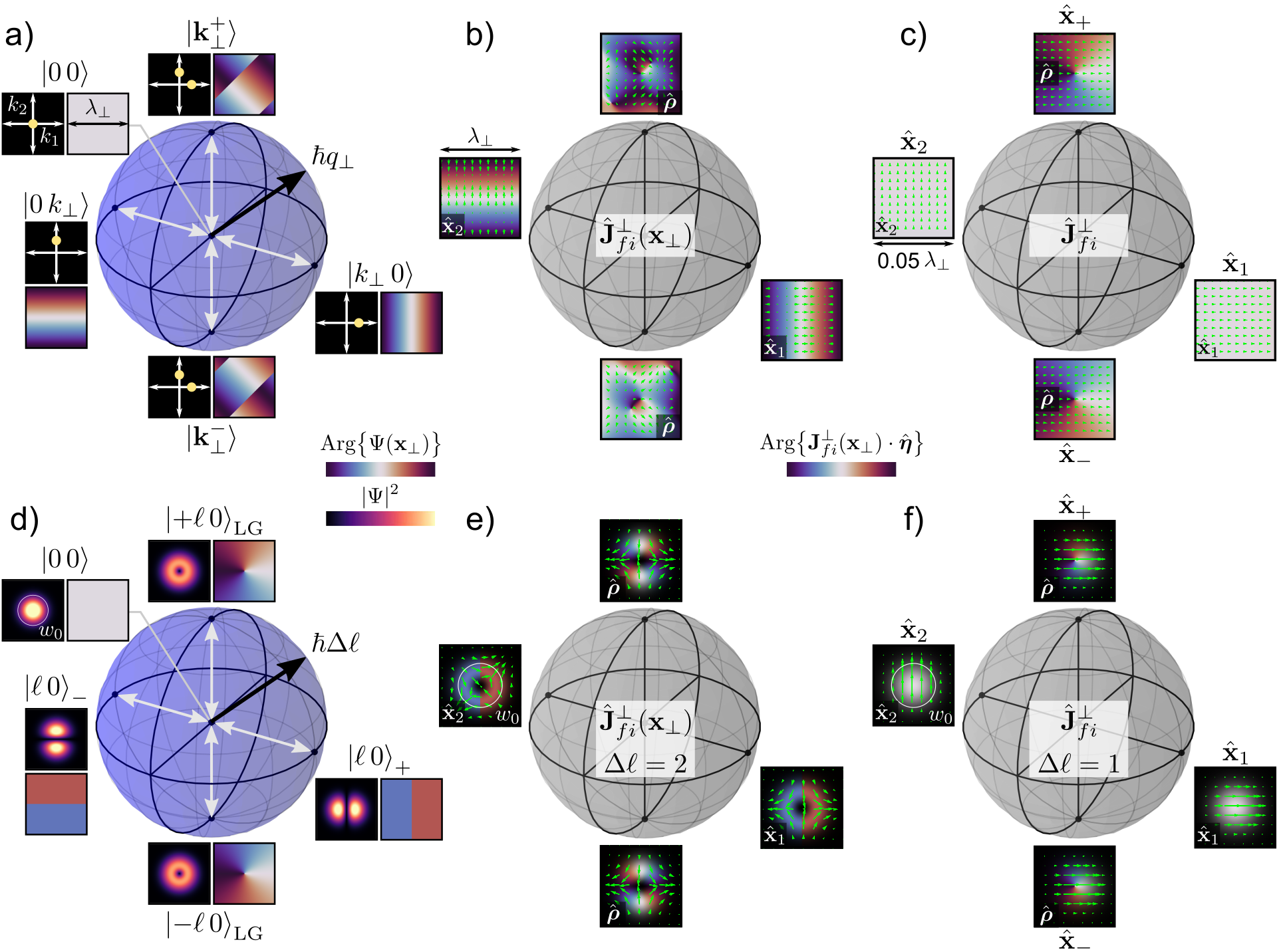}
    \caption{Optical polarization analogs arising from transitions between free electron states with well-defined linear and orbital angular momentum. a) Schematic representation of transitions between the $\ket{0\,0}$ transverse plane wave state with constant phase (sphere center) and other states with transverse wave vector magnitude $k_{\perp}$ and non-uniform phase arranged around the Poincar{\'e} sphere. Reciprocal space probability densities and real space transverse phase profiles over one transverse wavelength $\lambda_{\perp} = 2\pi/|\vec{q}_{\perp}|$ are shown for the states at the four antipodal states indicated. (b) Spatial maps of the transition current density $\vec{J}_{fi}(\vec{R}, z=0)$ (green arrows) for transverse state transitions represented by white arrows in panel (a). Underlying color maps represent the phase profiles of $\vec{J}^{\perp}_{fi}(\vec{x}_\perp) \cdot \hat{\boldsymbol{\eta}}$, where $\hat{\boldsymbol{\eta}}$ is noted within each plot. (c) Same as (b), but shown over a narrower region surrounding the origin with edge length $0.05\lambda_{\perp}$. (d) Schematic representation of transitions between the $\ket{0\,0}_{\textrm{LG}}$ state and superposition states with $\hbar \ell$ units of OAM arranged around the Poincar{\'e} sphere. Real space probability density and transverse phase profiles are shown for the states at the four antipodal states indicated for $\ell = 1$. The radius of the white circle is equal to the beam waist $w_0$. (e) $\vec{J}_{fi}(\vec{R}, z=0)$ (green arrows) for transverse state transitions represented by white arrows in panel (d) for $\Delta \ell = 2$. Underlying color maps represent the phase profiles of $\vec{J}^{\perp}_{fi}(\vec{x}_\perp) \cdot \hat{\boldsymbol{\eta}}$. (f) Same as (e), but for $\Delta \ell = 1$.
    }
    \label{F2}
\end{figure}

Consider, for example, transitions between electron plane wave states with $\ket{k_1\,k_2}$ denoting the transverse LM state with $\ket{0\,0}$ defining the plane wave oriented along the TEM axis. Figure \ref{F2}a shows a Poincar{\'e} sphere with the surface composed of all transverse plane wave states with fixed transverse wave vector magnitude $k_{\perp}$, and the $\ket{0\,0}$ state characterized by constant spatial phase profile at the origin. For any transitions between the $\ket{0\,0}$ state and a state on the sphere surface, the transverse linear momentum transfer $\hbar \vec{q}_{\perp} = \hbar \vec{k}_{\perp}$ fixes the radius of the sphere. Spanning the equatorial plane are the $\ket{k_{\perp}\,0}$ and $\ket{0\,k_{\perp}}$ states characterized by phase profiles independent of $x_2$ and $x_1$, respectively. The superposition states $\ket{\vec{k}^{\pm}_{\perp}} = (1/\sqrt{2})[ \ket{k_{\perp}\,0} \pm i \ket{0\,k_{\perp}}]$ are located at the vertical pair of antipodal points. The reciprocal space wave function density and real space transverse phase are shown over one transverse wavelength $\lambda_{\perp} = 2 \pi/q_{\perp}$ for the four antipodal points indicated.


Although $\ket{0\,0} \rightarrow \ket{0\,0}$ transitions lead to $\hat{\vec{J}}_{fi}(\vec{x})$ purely along $\hat{\bf x}_3$, the transverse transition current density arising from transitioning from $\ket{0\,0}$ to an arbitrary superposition of states on the surface of the sphere in Figure \ref{F2}a is $\vec{J}^{\perp}_{fi}(\vec{x}) = ({\hbar e}/{2 m L^2}) e^{i q_{\parallel}x_3} \int_{|{\bf k}_\perp|=q_\perp}\vec{k}_{\perp}d{\bf k}_{\perp}{\widetilde\Psi}_f^{*}(\vec{k}_{\perp}) e^{-i\vec{k}_{\perp} \cdot \vec{x}_{\perp}}/{(2\pi)^2},$ where ${\widetilde\Psi}_f(\vec{k}_{\perp})$ is the reciprocal space wave function of the final 2D transverse pure state and $L$ is the box quantization length. If the final state consists of a single plane wave component, $ \vec{J}_{fi}(\vec{x}) =(-{\hbar e}/{2mL^3}) (2\vec{k}_i - \vec{q}) e^{i \vec{q}\cdot \vec{x}}$, with $\vec{J}^{\perp}_{fi}(\vec{x}) \propto \vec{q}_\perp e^{i \vec{q}\cdot \vec{x}}$. Figure \ref{F2}b presents $\vec{J}^{\perp}_{fi}(\vec{x}_\perp, x_3=0)$ for transitions (marked as white arrows) between the $\ket{0\,0}$ state and the four antipodal states on the surface of the sphere in Figure \ref{F2}a. The spatial periodicity of the plane wave wave functions is inherited by $\vec{J}^{\perp}_{fi}(\vec{x}_\perp)$, leading to spatial variation of the sign of the $\hat{\vec{x}}_1$ and $\hat{\vec{x}}_2$ components of $\vec{J}^{\perp}_{fi}(\vec{x}_\perp)$ at the equatorial antipodal points, as well as sign and orientation variation at the vertically oriented antipodal points. However, when $|\vec{q}_{\perp}|d \ll 1$, where $d$ is the characteristic transverse length scale of the target, Figure \ref{F2}c shows $\hat{\vec{J}}^{\perp}_{fi}(\vec{x})$ becomes approximately independent of position and is oriented along $\hat{\mathbf{q}}_{\perp}$ in the vicinity of the target. Thus, the resulting $\hat{\vec{J}}^{\perp}_{fi} = \hat{\mathbf{q}}_{\perp}$ Poincar{\'e} sphere becomes equivalent to that used to characterize optical plane wave polarization states \cite{SM}. This is precisely the dipole limit  discussed previously in the core-loss literature identifying $\hat{\vec{q}}_{\perp}$ as the OPA in plane wave based measurements \cite{Hitchcock1993-vs, Yuan1997-hj, Hebert2003-mn, Schattschneider2006-fy}.

Another Poincar{\'e} sphere can be constructed using the cylindrically symmetric Laguerre-Gauss (LG) $\ket{\ell \, p}_{\textrm{LG}}$ transverse states (Figure \ref{F2}d), where $\hbar \ell$ is the OAM and $p$ labels the number of radial nodes. The $\ket{\pm\ell \, 0}_{\textrm{LG}}$ and superposition states $\ket{\ell\,0}_{+} = ({1}/{\sqrt{2}})\big( \ket{+\ell\,0}_{\textrm{LG}} + \ket{-\ell\,0}_{\textrm{LG}} \big)$ and $\ket{\ell\,0}_{-} = ({1}/{i\sqrt{2}})\big( \ket{+\ell\,0}_{\textrm{LG}} - \ket{-\ell\,0}_{\textrm{LG}} \big)$ are located at the vertical and equatorial antipodal points, respectively. As in Figure \ref{F2}a, the $\ket{0\,0}_{\textrm{LG}}$ state with constant transverse phase is positioned at the origin such that transitions between the state at the origin and states on the sphere surface are associated with transfers of $\Delta \ell$ units of OAM. The real space wave function densities and transverse phase profiles are presented for $\Delta \ell = 1$. $\vec{J}_{fi}(\vec{x})$ acquires components transverse to $\hat{\vec{x}}_3$ only when there is a transition between transverse free electron states. Figure \ref{F2}e shows $\vec{J}^{\perp}_{fi}(\vec{x}_{\perp})$ for transitions (marked as white arrows in Figure \ref{F2}d) between the uniform phase state at the origin and the four antipodal points shown on the surface for $\Delta \ell = 2$. While these transition current densities possess the symmetries required to excite quadrupolar target excitations, they do not constitute an OPA due to the spatial variation of $\hat{\vec{J}}_{fi}^{\perp}(\vec{x}_{\perp})$. This situation reflects the general relationship \cite{allen1992orbital} between the LG OAM states and the Hermite-Gauss (HG) states, the latter of which are separable in the Cartesian coordinate system. Only in the particular case of $\Delta \ell = 1$ are the equatorial antipodal states related to the first order HG states, characterized by phase profiles independent of $y$ and $x$, respectively, by $\ket{\ell =1 \,0}_{\pm} = \ket{ ^{1\,0}_{0\,1} }_{\textrm{HG}}$ \cite{lourencco2021optical}. In this case, the orientation and spatial phase profiles of $\hat{\vec{J}}_{fi}^{\perp}$ at the four antipodal points of Figure \ref{F2}f are identical to those associated with the electric field of circularly polarized light, satisfying the necessary OPA conditions.

Due to the delocalized (localized) nature of the plane wave (LG/HG) states, it is conventional to describe EEL measurements involving plane wave and LG/HG states in terms of the double differential scattering cross section ${\partial^2 \sigma}/{\partial E_{if} \partial \Omega}$ and the state- and energy-resolved EEL probability $\Gamma_{fi}$ observables, respectively. Specializing to the Cartesian coordinate system with $(x, y, z) = (\vec{R}, z)$ and impact parameter $\vec{R}=\vec{R}_0$, both observables are related to $w^{\textrm{loss}}_{fi}(\omega)$ in Eq. \eqref{w_fi_loss} by
\begin{equation}
    \left[
    \begin{array}{c}
        \frac{\partial^2 \sigma}{\partial E_{if}\partial \Omega} \\
        \Gamma_{fi}(\vec{R}_0, \omega)
        \end{array}
    \right]=
    \left[
    \begin{array}{c}
        L^6\Big(\frac{ 2m}{4\pi \hbar^2} \Big)^2 {\gamma} \Big(\frac{k_f}{k_i}\Big) \int \frac{d(\hbar\omega)}{2\pi} \\
        \frac{L^2}{\hbar v}\int\frac{dq_{\parallel}}{2\pi} 
    \end{array}
    \right]\times w^{\textrm{loss}}_{fi}(\omega),
\end{equation}
where $\gamma=1/\sqrt{1-(v/c)^2}$, and the non-recoil approximation $\delta(\omega - \omega_{if}) \approx (1/v)\delta(q_{\parallel} - \omega/v)$ is invoked in the lower expression \cite{SM}. These observables are compared for the representative nanophotonic system composed of two 60 nm $\times$ 30 nm $\times$ 15 nm Ag rods with a 10 nm surface-to-surface gap along the dimer ($\hat{\mathbf{y}}$) axis.

\begin{figure}
     \centering
     \includegraphics{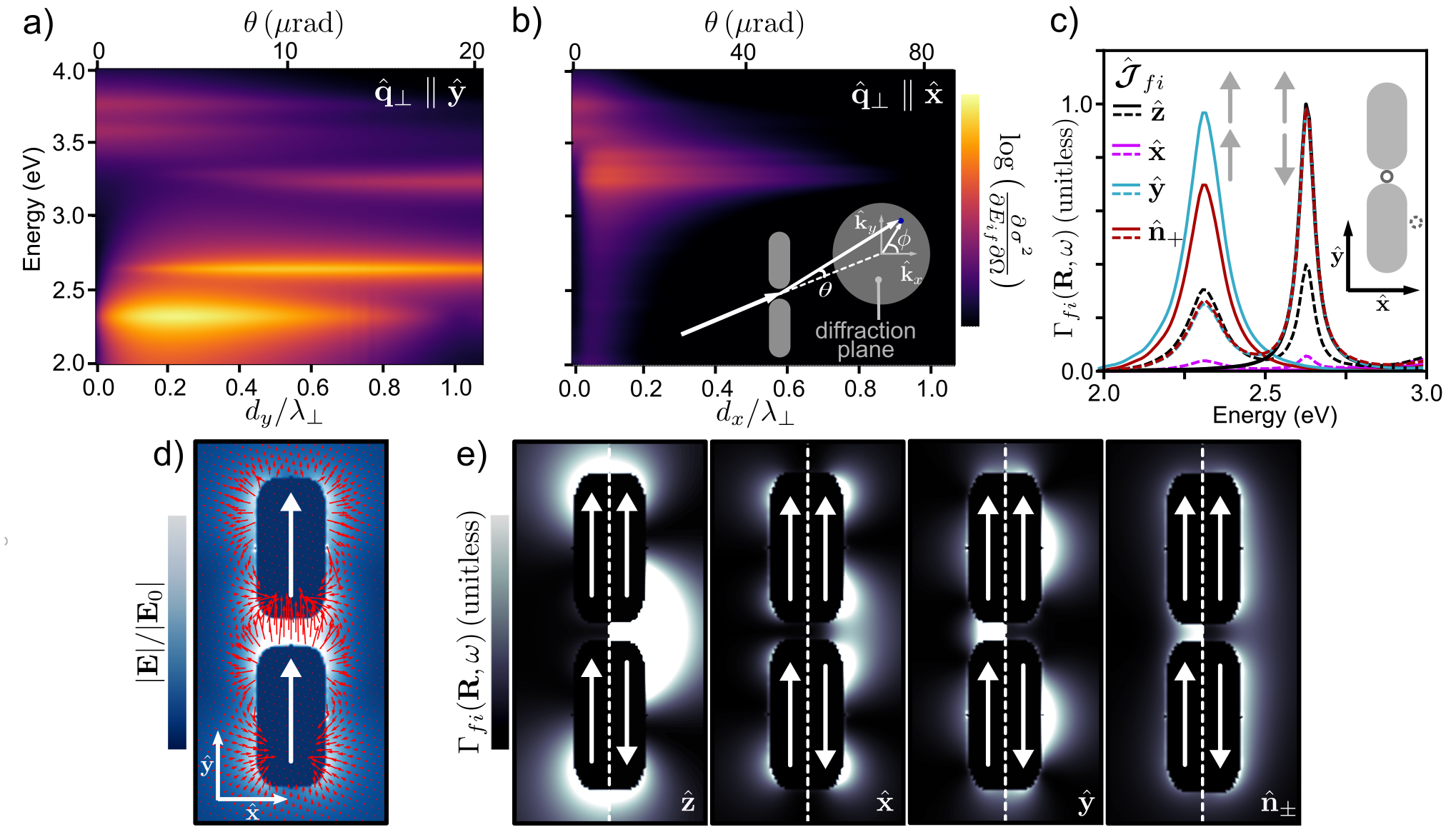}
     \caption{Phase-shaped EEL measurements of an Ag rod dimer system. (a) Double differential inelastic scattering cross section in the low-loss spectral region for scattering angles $(\theta, \phi) = (0-20$ $\mu$rad, $\pi/2)$ such that $\mathbf{q}_{\perp} = q_{\perp} \hat{\mathbf{y}}$. (b) Same as (a) but for scattering angles $(\theta, \phi) = (0-80$ $\mu$rad, $0)$ such that $\mathbf{q}_{\perp} = q_{\perp} \hat{\mathbf{x}}$. (c) $\Gamma_{fi}(\mathbf{R}, \omega)$ within the spectral region containing the bonding (2.31 eV) and antibonding (2.64 eV) hybridized dipole modes for $\mathbf{R}_0$ on (solid) and off (dashed) the center of mass position. Trace colors denote $\hat{\boldsymbol{\mathcal{J}}}_{fi}$. (d) Optically-induced response field at the bonding mode energy for incident wave vector along $\hat{\mathbf{z}}$ and polarization along $\hat{\mathbf{y}}$. In-plane $xy$ components of the vector field are shown in red. (e) Polarization-resolved spectrum images of bonding (left) and antibonding (right) modes for transitions between first order LG/HG states and the Gaussian state.  $\hat{\boldsymbol{\mathcal{J}}}_{fi}$ is labeled in each spectrum image.}
     \label{F3}
 \end{figure}

Figures \ref{F3}a,b show normalized ${\partial^2 \sigma}/{\partial E_{if} \partial \Omega}$ spectra (log scale) for 200 keV electrons in the loss energy window containing the rods' coupled surface plasmon modes, which were computed using the $e$-DDA code \cite{Bigelow2012-pq, bigelow2013signatures, SM}. In Figure \ref{F3}a, the incoming electron plane wave is aligned along the TEM axis $(\theta,\phi)=(0,0)$ and the outgoing plane waves emerge at angles $(\theta,\phi)=(0-20\ \mu{\textrm{rad}},\pi/2)$ such that $\hat{\mathbf{q}}_{\perp}$ is purely along $\hat{\mathbf{y}}$. In this low-loss regime, the dipole limit with $d_y \lesssim 0.05 \lambda_{\perp}$ (Figure \ref{F2}c) is achieved for $\theta \lesssim 1$ $\mu$rad. As expected, the spectrum in Figure \ref{F3}a for $\theta = 1$ $\mu$rad is dominated by the bonding dipole mode at 2.31 eV, which is accessible by optical plane wave excitation polarized along $\hat{\mathbf{q}}_{\perp} = \hat{\mathbf{y}}$ \cite{SM}. As the detection angle increases, $\lambda_{\perp}$ decreases and higher order excitations such as the $\hat{\mathbf{y}}$ oriented antibonding mode at 2.64 eV grow into the spectra \cite{soininen2005inelastic}. Figure \ref{F3}b presents ${\partial^2 \sigma}/{\partial E_{if} \partial \Omega}$ spectra for detection angles $(\theta,\phi)=(0-80\ \mu{\textrm{rad}},0)$, such that $\hat{\mathbf{q}}_{\perp}$ is along $\hat{\mathbf{x}}$. Besides the longitudinally oriented modes above 3.5 eV that dominate at $\theta = 0 $ in panels (a) and (b), the modes near 3.25 eV in the optical extinction spectra under $\hat{\mathbf{x}}$ polarization appear for $\mathbf{q}_x \ne 0$ \cite{SM}. Clear connection between optical $\sigma_{\textrm{ext}}(\omega)$ and ${\partial^2 \sigma}/{\partial E_{if} \partial \Omega}$ exists at small detection angles consistent with the dipole limit (Figure \ref{F2}c). The required angular resolution ($\lesssim 1$ $\mu$rad in the present case) is currently achievable experimentally \cite{shekhar2017momentum, krehl2018spectral, saito2019emergence, PhysRevResearch.2.043227}.

In the case of localized LG state transitions (Figure \ref{F2}d), $\Gamma_{fi}$ can be put into the form 
\begin{equation}
    \begin{aligned}
        \Gamma_{fi}(\vec{R}_0, \omega) = \frac{1}{\pi \hbar^2 \omega}\textrm{Re}\bigg\{ - \int d\mathbf{R} dz ~ \bigg( \frac{L}{v}\bigg)\vec{J}^*_{fi}(\vec{R}, z) \cdot \vec{E}_{fi}(\vec{R}, z, \omega) \bigg\}
    \end{aligned}
    \label{Gamma_E_tr}
\end{equation}
by introducing the induced field $\vec{E}_{fi}(\vec{x}, \omega ) = -4 \pi i \omega \int d\mathbf{x}' ~\tensor{\mathbf{G}}(\vec{x}, \vec{x}', \omega) \cdot  (L/v)\vec{J}_{fi}(\vec{x}')$ sourced by the transition current density in the presence of the target, and $\omega = q_{\parallel} v$ within the non-recoil approximation. When $w_0$ is small such that $\vec{E}_{fi}(\vec{x}, \omega) \approx \vec{E}_{fi}(\vec{R}_0, z, \omega)$,  
\begin{equation}
    \Gamma_{fi}(\vec{R}_0, \omega) \approx \frac{1}{\pi \hbar^2 \omega}\textrm{Re}\bigg\{ - \Big(\frac{L}{v}\Big)|\mathcal{J}_{fi}|\int dz ~\hat{\boldsymbol{\mathcal{J}}}^*_{fi}e^{-i \omega z/v} \cdot \vec{E}_{fi}(\vec{R}_0, z, \omega)  \bigg\},
    \label{Gamma_fi_smallw}
\end{equation}
where $\boldsymbol{\mathcal{J}}_{fi} = \int d\mathbf{R}\boldsymbol{\mathbf{J}}_{fi}(\mathbf{R}) e^{-i \omega z/v}= ({i \hbar e}/{2m})\{ \big[ \braket{\Psi_f | \nabla_{\perp}|\Psi_i} - \braket{\Psi_i | \nabla_{\perp}|\Psi_f}^* \big] + i (2 k_i - q_\parallel)  \braket{\Psi_f | \Psi_i} \hat{\vec{z}}\}$.
In this limit, $\hat{\boldsymbol{\mathcal{J}}}^{\perp}_{fi} = \mathbf{0}$ unless $\Delta \ell = \pm1$ in which case $\hat{\boldsymbol{\mathcal{J}}}^{\perp}_{fi} = \hat{\vec{J}}_{fi}^{\perp} = \hat{\bf d}^{\perp}_{fi}$ \cite{SM} as shown in Figure \ref{F2}f. Figure \ref{F3}c presents $\Gamma_{fi}(\mathbf{R}, \omega)$ evaluated numerically with the $e$-DDA code \cite{Bigelow2012-pq, bigelow2013signatures, draine2008discrete, SM} using Eq. \eqref{Gamma_fi_smallw} for $\hat{\boldsymbol{\mathcal{J}}}^{\perp}_{fi}$ at the antipodal points in Figure \ref{F2}f as well as $\hat{\boldsymbol{\mathcal{J}}}^{\perp}_{fi} \parallel \hat{\mathbf{z}}$ for the transition $\ket{0\,0}_{\textrm{LG}} \rightarrow \ket{0\,0}_{\textrm{LG}}$ within the spectral region containing the bonding and antibonding hybridized dipole modes. Trace colors denote $\hat{\boldsymbol{\mathcal{J}}}_{fi}$, while solid and dashed lines indicate $\mathbf{R}_0$ positioned at, or displaced from, respectively, the dimer center of mass. At the center of mass position, a conventional EEL process with no transition between transverse state (solid black) does (does not) couple to the optically dark (bright) antibonding (bonding) mode. In contrast, when $\hat{\boldsymbol{\mathcal{J}}}_{fi} = \hat{\boldsymbol{\mathcal{J}}}^{\perp}_{fi}$, the loss functions exhibit peaks at the bonding, but not antibonding, mode. When $\mathbf{R}_0$ is displaced to a position of lower symmetry, all electron scattering processes considered couple to both bonding and antibonding modes. 

Polarization-resolved spectrum images can be collected via hyperspectral imaging by scanning $\mathbf{R}_0$ over the target specimen. Although $\Gamma_{fi}(\mathbf{R}, \omega)$ is nonlocal in the $z$-direction \cite{de2008probing, hohenester2009electron}, conventional ($\hat{\boldsymbol{\mathcal{J}}}_{fi} = \hat{\mathbf{z}}$) and polarization-resolved ($\hat{\boldsymbol{\mathcal{J}}}_{fi} = \hat{\boldsymbol{\mathcal{J}}}_{fi}^{\perp}$) spectrum images can often be rationalized by considering the electric field (Figure \ref{F3}d). Figure \ref{F3}e presents polarization-resolved spectrum images for loss energies matching the bonding (left) and antibonding (right) mode energy, and for transitions between electron transverse states indicated by the transition current unit vector $\hat{\boldsymbol{\mathcal{J}}}_{fi}$ labeled in each image. It is evident upon comparison of the spectrum images at the bonding mode energy and Figure \ref{F3}d that the regions of space where $\Gamma_{fi}(\mathbf{R},\omega)$ is large closely track positions where $\mathbf{E} \cdot \hat{\boldsymbol{\mathcal{J}}}_{fi}$ is appreciable. The polarization-resolved spectrum images at the antibonding energy exhibit the expected nodal behavior at the origin for each $\hat{\boldsymbol{\mathcal{J}}}_{fi}$. This example demonstrates the additional wealth of information that can be accessed using OPAs in phase-shaped EEL measurements, highlighting commonalities and differences arising for pre- and post-selection of transverse plane wave and LG/HG states.

Owing to recently developed methods for manipulating free electron wave functions, inelastic electron scattering between selected phase-shaped transverse states constitutes a powerful addition to the rapidly-developing nanoscale imaging toolset. By employing a quantum mechanical treatment that explicitly accounts for the transverse electron degrees of freedom with fully-retarded light-matter interactions, we show that the transition current density ${\mathbf{J}}^{\perp}_{fi}$ plays the role of OPA in EEL measurements and provide a general prescription for constructing OPAs that mimic free space optical plane waves. Specifically, we demonstrate an underlying connection between the two ostensibly distinct OPAs previously identified in the core- and low-loss regimes under wide-field and focused beam conditions and discuss the conditions required to closely approximate ideal OPAs. Example calculations for a plasmonic rod dimer are presented to highlight the utility of phase-shaped EEL nanospectroscopy for determining mode symmetries and probing the 3D polarization-resolved response field of a target with nanoscale spatial resolution. Although primary focus is placed on developing free space photon OPAs, the procedure outlined for constructing $\hat{{\mathbf{J}}}^{\perp}_{fi}$ along arbitrary curvilinear coordinate directions is general and can be applied to generate more exotic structured light \cite{forbes2021structured} analogs such as azimuthally $\hat{{\mathbf{J}}}^{\perp}_{fi} = \hat{\boldsymbol{\phi}}$ and radially $\hat{{\mathbf{J}}}^{\perp}_{fi} = \hat{\boldsymbol{\rho}}$ polarized transition current densities that couple to other desired target mode symmetries. The fully-retarded formalism employed here is consistent with that used to describe laser-stimulated phase-shaped electron energy gain measurements \cite{bourgeois2022polarization}, setting the stage for time-resolved phase-shaped measurements in ultrafast TEMs \cite{PhysRevLett.126.233403, garcia2021optical}, which complement other methods with state-of-the-art time and space resolution such as interferometric time-resolved photoemission electron microscopy \cite{dai2020plasmonic}.

\begin{acknowledgments}
All work was supported by the U.S. Department of Energy (DOE), Office of Science, Office of Basic Energy Sciences (BES), Materials Sciences and Engineering Division under Award No. DOE BES DE-SC0022921. 
\end{acknowledgments}


\bibliography{refs}
\end{document}


\author{Marc R. Bourgeois}
\affiliation
{Department of Chemistry, University of Washington, Seattle WA, 98195}
\author{Austin G. Nixon}
\affiliation
{Department of Chemistry, University of Washington, Seattle WA, 98195}
\author{Matthieu Chalifour}
\affiliation
{Department of Physics, University of Washington, Seattle WA, 98195}
\author{David J. Masiello}
\email{masiello@uw.edu}
\affiliation
{Department of Chemistry, University of Washington, Seattle WA, 98195}

\title{Supplemental Material:\\Optical Polarization Analogs in Inelastic Free Electron Scattering}



\maketitle
\tableofcontents

\newpage
\section{Derivation of the State- and Frequency-Resolved Electron Energy Loss Rate}\label{section_Derivation_MTEq1} 
The interaction between a swift electron and a target consisting of dielectric material characterized by $\varepsilon (\vec{x}, \omega)$ is evaluated using time-dependent scattering theory. Working in the generalized Coulomb Gauge \cite{glauber1991quantum} defined by $\nabla \cdot \varepsilon(\vec{x})\vec{A}(\vec{x},t) = 0$, the interaction potential is
\begin{equation}
    \hat{V} =\frac{e}{2mc} \bigg(\hat{\vec{A}} \cdot \hat{\vec{p}} + \hat{\vec{p}} \cdot \hat{\vec{A}} \bigg),
    \label{V_interaction}
\end{equation}
where $m$ is the mass of the electron, $e$ is the electron charge, $c$ is the speed of light, and $\hat{\vec{A}}$ and $\hat{\vec{p}}$ represent the electromagnetic vector potential of the target system and the canonical momentum of the free electron, respectively. Within the first Born approximation, the rate at which a free electron prepared in initial state $| i \rangle $ transitions to a final state $| f \rangle $, thereby adding $n$ quanta of energy into the $\ell^{\textrm{th}}$ target mode ( $| 0_{\ell} \rangle \rightarrow | n_{\ell} \rangle $), is given by Fermi's golden rule (units of s$^{-1}$)
\begin{equation}
    w^{\textrm{loss}}_{fi} = \frac{2 \pi}{\hbar} \big| \langle f | \langle n_{\ell}| \hat{V} | 0_{\ell} \rangle |i \rangle \big|^{2} \delta(E_f^{\textrm{total}} - E_i^{\textrm{total}}).
    \label{w_fi_loss}
\end{equation}
The frequency-resolved transition rate is obtained by writing the energy conserving delta function in the form
$ \delta(E_f^{\textrm{total}} - E_i^{\textrm{total}}) = \frac{1}{\hbar} \int d \omega \, \delta(\omega - \omega_{if}) \delta( \omega + \omega_{\ell})$, where $\hbar \omega_{if}$ is the energy loss of the free electron and $\hbar \omega_{\ell}$ is the energy of the excited target mode. Consequently (now unitless),    \begin{equation}
    w^{\textrm{loss}}_{fi} (\omega ) = \frac{2 \pi}{\hbar^2} \big| \langle f | \langle n_{\ell}| \hat{V} | 0_{\ell} \rangle |i \rangle \big|^{2} \delta (\omega - \omega_{if}) \delta(\omega + \omega_{\ell}).
    \label{w_fi_loss_resolved}
\end{equation}
Now we can start evaluating the matrix elements for an arbitrary electron transition ($|i \rangle \rightarrow | f \rangle$) due to its interaction with a target system. Inserting two resolutions of the identity ($1 = \int d \vec{x} | \vec{x} \rangle \langle \vec{x} | $) into initial and final bra-ket states and carrying out the necessary integration, we arrive at the familiar spatial integration over $\hat{\vec{A}}(\vec{x}) \cdot \vec{J}_{fi}(\vec{x})$ (with units of erg)
\begin{equation}
    \begin{aligned}
        \langle f | \hat{V} | i \rangle &=  \frac{e}{2mc} \int d\vec{x} \, d\vec{x}' \, \langle f | \vec{x}' \rangle \langle \vec{x}'|\hat{\vec{A}} \cdot \hat{\vec{p}} + \hat{\vec{p}} \cdot \hat{\vec{A}} | \vec{x} \rangle \langle \vec{x} | i \rangle \\
        &=\frac{-i\hbar e}{2mc} \int d\vec{x} \, \hat{\vec{A}}(\vec{x}) \cdot \psi_{f}^{*}(\vec{x}) \nabla \psi_{i}(\vec{x}) - \frac{i\hbar e}{2mc} \int d\vec{x} \, \Big( \psi_{f}^{*}(\vec{x}) \psi_{i}(\vec{x}) \big[ \nabla \cdot \hat{\vec{A}}(\vec{x}) \big] + \hat{\vec{A}}(\vec{x}) \cdot  \psi_{f}^{*}(\vec{x}) \nabla \psi_{i}(\vec{x})\Big) \\
        &=\frac{-2i\hbar e}{2mc} \int d\vec{x} \, \hat{\vec{A}}(\vec{x}) \cdot \psi_{f}^{*}(\vec{x}) \nabla \psi_{i}(\vec{x}) - \frac{i\hbar e}{2mc} \int d\vec{x} \, \psi_{f}^{*}(\vec{x}) \psi_{i}(\vec{x}) \big[ \nabla \cdot \hat{\vec{A}}(\vec{x}) \big] \\
        & = \frac{-2i\hbar e}{2mc} \int d\vec{x} \, \hat{\vec{A}}(\vec{x}) \cdot \psi_{f}^{*}(\vec{x}) \nabla \psi_{i}(\vec{x}) + \frac{i\hbar e}{2mc} \int d\vec{x} \, \hat{\vec{A}}(\vec{x}) \cdot \Big\{\nabla \psi_{f}^{*}(\vec{x}) \psi_{i}(\vec{x})+\psi_{f}^{*}(\vec{x}) \nabla \psi_{i}(\vec{x}) \Big\} \\
        & = \frac{1}{c} \int d\vec{x} \, \hat{\vec{A}}(\vec{x}) \cdot \frac{-i\hbar e}{2m} \Big\{ \psi_{f}^{*}(\vec{x}) \nabla \psi_{i}(\vec{x}) - \psi_{i}(\vec{x})\nabla \psi_{f}^{*}(\vec{x}) \Big\} \\
        & = -\frac{1}{c} \int d\vec{x} \, \hat{\vec{A}}(\vec{x}) \cdot \vec{J}_{fi}(\vec{x}).
    \end{aligned}
    \label{matrix_el_fi}
\end{equation}
Integration by parts was used in order to go from the third to fourth lines of Eq. \eqref{matrix_el_fi} above, importantly noting that $\int d \vec{x} \, \psi^*_f(\vec{x}) \psi_i(\vec{x}) \big[ \nabla \cdot \vec{A}(\vec{x}) \big] = - \int d \vec{x} \, \bigg\{ \big[\nabla \psi^*_f(\vec{x}) \big] \psi_i(\vec{x}) + \psi^*_f(\vec{x}) \big[\nabla \psi_i(\vec{x}) \big] \bigg\} \cdot \vec{A}(\vec{x})$, arising from the surface term being equal to zero, i.e, $\int d \vec{x} \, u(\vec{x}) \big[ \nabla \cdot \vec{V} (\vec{x}) \big] = - \int d \vec{x} \, \nabla u(\vec{x}) \cdot \vec{V}(\vec{x})$. In the last lines of Eq. \eqref{matrix_el_fi}, we explicitly arrive at an expression for the effective transition current density defined as
\begin{equation}
    \vec{J}_{fi}(\vec{x}) = \frac{i \hbar e}{2m} \bigg\{ \psi^*_f(\vec{x}) \nabla \psi_i(\vec{x})  - \psi_i(\vec{x}) \nabla \psi^*_f(\vec{x}) \bigg\}, 
    \label{J_fi}
\end{equation}
in statC$\cdot$s$^{-1}$$\cdot$cm$^{-2}$. Up to this point, we have remained agnostic as to the exact nature of both the target system and electron initial and final wavefunctions. In order to arrive at a total transition rate for the electron to lose $\hbar \omega_{if}$, we sum over all the possible final states of the target, $\ell$ in Eq. \eqref{w_fi_loss_resolved}. We can expand the vector potential as $\vec{A}(\vec{x}, t) = \, \sum_{n}\hat{a}^{\dagger}_{n} \vec{A}_{n}^{(-)}(\vec{x}) e^{+i\omega_{n} t} +  \hat{a}_{n} \vec{A}_{n}^{(+)}(\vec{x}) e^{-i\omega_{n} t} $, where $\vec{A}_{n}^{(+)}(\vec{x}) = c\sqrt{\frac{2\pi \hbar}{\omega_n}} \, \vec{f}_{n}(\vec{x})$ and $\vec{A}_{n}^{(-)}(\vec{x}) = c\sqrt{\frac{2\pi \hbar}{\omega_n}} \, \vec{f}_{n}^{*}(\vec{x})$ are the positive and negative frequency components of the vector potential written in terms of spatial mode functions $\vec{f}_{n}(\vec{x})$, which satisfy the generalized Helmholtz equation. Since we are considering only electron energy loss, $\sum_{\ell} \langle 1_{\ell}\big| \hat{\vec{A}}(\vec{x}',t)\big| 0_{\ell} \rangle = \sum_{n, \ell} c \sqrt{\frac{2\pi \hbar}{\omega_{\ell}}} \vec{f}_{\ell }^*(\vec{x}') e^{i \omega_{\ell} t} \delta_{n, \ell } = \sum_{n} \vec{A}_{n}^{(-)}(\vec{x}')$  and consequently for the conjugate process, $\sum_{\ell} \langle 0_{\ell }\big| \hat{\vec{A}}(\vec{x},t)\big| 1_{\ell} \rangle^*  = \sum_{n, \ell} c \sqrt{\frac{2\pi \hbar}{\omega_{\ell}}} \vec{f}_{\ell}(\vec{x}) e^{-i \omega_{\ell} t} \delta_{n, \ell} = \sum_{n} \vec{A}_{n}^{(+)}(\vec{x})$. Inserting these matrix elements into Eq. \eqref{w_fi_loss_resolved} one obtains Eq. (1) in the main text (the frequency-resolved rate is unitless)
\begin{equation}
    \begin{aligned}
        w^{\textrm{loss}}_{fi} (\omega ) & = \sum_{\ell} \frac{2 \pi}{c^2 \hbar^2} \big|\langle 1_{\ell}\big| \int d\vec{x} \, \hat{\vec{A}}(\vec{x}) \cdot \vec{J}_{fi}(\vec{x}) \big| 0_{\ell} \rangle \big|^{2} \delta (\omega - \omega_{if}) \delta(\omega + \omega_{\ell}) \\
        & = \sum_{n} \frac{2 \pi}{c^2 \hbar^2} \int d\vec{x} \, d\vec{x}' \, \big[ \vec{A}_{n}^{(+)}(\vec{x}) \cdot \vec{J}_{fi}^*(\vec{x})\big]  \big[ \vec{A}_{n}^{(-)}(\vec{x}) \cdot \vec{J}_{fi}(\vec{x}')\big] \delta (\omega - \omega_{if}) \delta(\omega + \omega_{n}) \\
        & = \frac{4 \pi^2}{\hbar} \int d\vec{x} \, d\vec{x}' \, \vec{J}_{fi}^*(\vec{x}) \cdot \bigg[ \sum_{n}\frac{1}{\omega_n} \vec{f}_{n}(\vec{x}) \vec{f}_{n}^*(\vec{x}') \delta(\omega + \omega_{n})\bigg]  \cdot\vec{J}_{fi}(\vec{x}') \delta (\omega - \omega_{if}) \\
        & = \frac{4 \pi^2}{\hbar} \int d\vec{x} \, d\vec{x}' \, \vec{J}_{fi}^*(\vec{x}) \cdot \bigg[ -\frac{2}{\pi} \textrm{Im}\bigg\{ \tensor{\mathbf{G}}(\vec{x}, \vec{x}', \omega) \bigg\}\bigg]  \cdot\vec{J}_{fi}(\vec{x}') \delta (\omega - \omega_{if}) \\
        & = \frac{4 \pi^2}{\hbar \omega} \int d\vec{x} \, d\vec{x}' \, \vec{J}_{fi}^*(\vec{x}) \cdot  \tensor{\boldsymbol{\varrho}}(\vec{x}, \vec{x}', \omega) \cdot\vec{J}_{fi}(\vec{x}') \delta ( \omega - \omega_{if}).
    \end{aligned}
    \label{w_fi_loss_Eq1}
\end{equation}
Due to the reciprocity property of the Green dyadic $\tensor{ \mathbf{G}}(\vec{x}_1, \vec{x}_2, \omega) = \tensor{\mathbf{G}^{T}}(\vec{x}_2, \vec{x}_1, \omega)$,
\begin{equation}
    \int d\mathbf{x} \, d\mathbf{x}'\, \vec{J}^{*}_{fi}(\vec{x}) \cdot \textrm{Im}\bigg\{ \tensor{\mathbf{G}}(\vec{x}, \vec{x}', \omega) \bigg\}
  \cdot \vec{J}_{fi}(\vec{x}') = \int d\mathbf{x} \, d\mathbf{x}'\, \textrm{Im}\big\{ \vec{J}^{*}_{fi}(\vec{x}) \cdot\tensor{\bf G}(\vec{x}, \vec{x}', \omega) \cdot \vec{J}_{fi}(\vec{x}')\big\},
  \label{G_property}
\end{equation} 
and Eq. \eqref{w_fi_loss_Eq1} can be expressed as
\begin{equation}
    \begin{aligned}
        w^{\textrm{loss}}_{fi} (\omega ) & = \frac{4 \pi^2}{\hbar} \int d\vec{x} \, d\vec{x}' \, \vec{J}_{fi}^*(\vec{x}) \cdot \bigg[ -\frac{2}{\pi} \textrm{Im}\bigg\{ \tensor{\mathbf{G}}(\vec{x}, \vec{x}', \omega) \bigg\}\bigg]  \cdot\vec{J}_{fi}(\vec{x}') \delta (\omega - \omega_{if}) \\
        & = -\frac{8 \pi}{\hbar} \int d\vec{x} \, d\vec{x}' \, \vec{J}_{fi}^*(\vec{x}) \cdot \textrm{Im}\bigg\{ \tensor{\mathbf{G}}(\vec{x}, \vec{x}', \omega) \bigg\}  \cdot\vec{J}_{fi}(\vec{x}') \delta (\omega - \omega_{if}) \\
        & = -\frac{8 \pi}{\hbar} \int d\vec{x} \, d\vec{x}' \, \textrm{Im}\bigg\{ \vec{J}_{fi}^*(\vec{x}) \cdot \tensor{\mathbf{G}}(\vec{x}, \vec{x}', \omega)  \cdot\vec{J}_{fi}(\vec{x}') \bigg\}\delta (\omega - \omega_{if}).
    \end{aligned}
    \label{w_fi_loss_res_altform}
\end{equation}

It is clear from Eq. \eqref{J_fi} that interchanging the initial and final states $i \smallleftrightarrow f$ amounts to replacing $\vec{J}_{fi}(\vec{x}) \leftrightarrow \vec{J}_{fi}^*(\vec{x})$, which establishes the detailed balancing condition $w_{i \rightarrow f}^{\textrm{loss}} = w_{f \rightarrow i}^{\textrm{gain}}$.


\section{Free Space Optical Plane Waves}\label{section_optical_planewaves}

\begin{figure}
    \centering
    \includegraphics{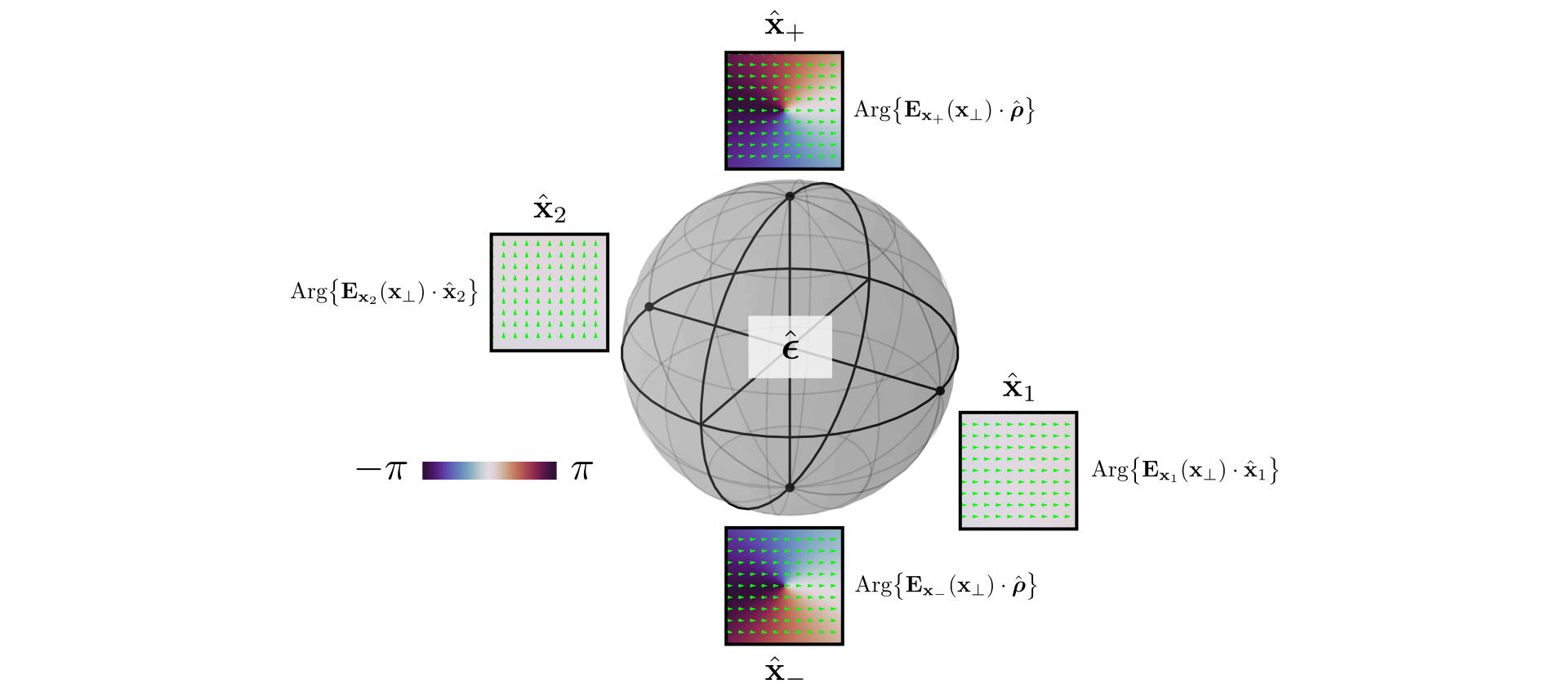}
    \caption{Free space optical polarization states $\hat{\boldsymbol{\epsilon}}$ represented by the Poincar{\'e} sphere.}
    \label{FS1}
\end{figure}

An arbitrary free photon pure polarization state is described by a point on the Poincar{\'e} sphere with pairs of antipodal points $\{\hat{\boldsymbol{\epsilon}}_{1},\hat{\boldsymbol{\epsilon}}_{2}\}$ and $\{\hat{\boldsymbol{\epsilon}}_{+},\hat{\boldsymbol{\epsilon}}_{-}\}$ that span the two-dimensional photon polarization Hilbert space (Figure S\ref{FS1}). The two sets of basis vectors are related by
\begin{equation}
    \begin{bmatrix}
        \hat{\bm\epsilon}_+ \\
        \hat{\bm\epsilon}_-
    \end{bmatrix} = \frac{1}{\sqrt{2}}\begin{bmatrix}
        1 & -i\\
        1 & i 
    \end{bmatrix} \begin{bmatrix}
        \hat{\bm\epsilon}_1 \\
        \hat{\bm\epsilon}_2
    \end{bmatrix}.
\end{equation}
Spatial maps of the electric field $\vec{E}(\mathbf{x}_{\perp}, x_3=0)$ of an optical plane wave with wave vector along $\hat{\mathbf{x}}_3$ (green arrows) are presented for each of the four antipodal points shown.  Underlying color maps represent the transverse phase profiles $\vec{E}_{\hat{\boldsymbol{\epsilon}}}(\mathbf{x}_{\perp}, x_3=0) \cdot \hat{\boldsymbol{\eta}}$, where $\hat{\boldsymbol{\epsilon}}$ is the unit vector describing the polarization state of the electric field, and $\hat{\boldsymbol{\eta}}$ is a unit vector within the $\mathbf{x}_{\perp}$ plane. At the vertical antipodal points $\hat{\mathbf{x}}_{\pm}$, for example, the phase of the  $\hat{\boldsymbol{\eta}} = \hat{\boldsymbol{\rho}}$ component of the electric field is plotted, where $\hat{\boldsymbol{\rho}}$ is the radial unit vector in polar coordinates. 

\subsection{Optical Extinction Cross Section}\label{ssection_FSOPW_cross_section}

The optical plane wave extinction cross section for an isolated dipolar target is
\begin{equation}
    \begin{aligned}
        \sigma_{\textrm{ext}}(\omega) = 4 \pi \frac{\omega}{c} \textrm{Im} \big\{ \hat{\boldsymbol{\epsilon}^{*}} \cdot  \tensor{\boldsymbol{\alpha}}(\omega) \cdot \hat{\boldsymbol{\epsilon}} \big\}, \\
    \end{aligned} 
\end{equation}
where $\tensor{\bm\alpha}( \omega)$ is the dipole polarizability tensor characterizing the target response to plane wave excitation with polarization unit vector $\hat{\boldsymbol{\epsilon}}$. This expression can be rewritten as 
\begin{equation}
    \begin{aligned}
        \sigma_{\textrm{ext}}(\omega) = 4 \pi \frac{\omega}{c} \frac{1}{|\mathbf{E}(\mathbf{x}_t)|^2} \textrm{Im} \bigg\{ \mathbf{E}^{*}(\mathbf{x}_t) \cdot  \tensor{\boldsymbol{\alpha}}(\omega) \cdot \mathbf{E}(\mathbf{x}_t) \bigg\}, \\
    \end{aligned} 
\end{equation}
where $\mathbf{E}(\mathbf{x}_t)$ is the plane wave electric field at the position of the target $\mathbf{x}_{t}$. A plane wave with wave vector along $\hat{\mathbf{k}}$ and polarization $\hat{\boldsymbol{\epsilon}}$ can be viewed as though sourced by a point dipole current density $\mathbf{J}(\mathbf{x}) = -i\omega p \delta(\mathbf{x} - \mathbf{x}_{\infty}) \hat{\boldsymbol{\epsilon}}$, since \cite{garcia2021optical} 
\begin{equation}
    \begin{split}
        \mathbf{E}(\mathbf{x}) &= -4 \pi i \omega \int d\mathbf{x}' \tensor{\bf G}_0(\mathbf{x}, \mathbf{x}', \omega) \cdot \mathbf{J}(\mathbf{x}') \\
        &= \bigg[ k^2\tensor{\bf I} + \nabla \nabla \bigg] \frac{e^{i k |\vec{x} - \vec{x}_\infty|}}{|\vec{x} - \vec{x}_\infty|} \cdot p \hat{\boldsymbol{\epsilon}},
    \end{split}
    \label{E_free_field}
\end{equation}
where
\begin{equation}
    \tensor{\bf G}_0(\mathbf{x}, \mathbf{x}', \omega) = -\frac{1}{4\pi \omega^2} \bigg[ k^2\tensor{\bf I} + \nabla \nabla \bigg] \frac{e^{i k |\vec{x} - \vec{x}'|}}{|\vec{x} - \vec{x}'|}
    \label{G0_dyadic_def}
\end{equation}
is the free space Green dyadic with units of s$^2$$\cdot$cm$^{-3}$, and both  $\mathbf{J}$ and $\mathbf{E}$ are understood to oscillate harmonically as $e^{-i \omega t}$. Because the source dipole is positioned at $\mathbf{x}_{\infty}$, 
\begin{equation}
    \mathbf{E}(\mathbf{x}) \to \frac{i}{ \omega} \int d\mathbf{x}'  k^2  \frac{e^{i k |\vec{x} - \vec{x}'|}}{|\vec{x} - \vec{x}'|} \tensor{\bf I} \cdot \mathbf{J}(\mathbf{x}')
    \label{E_free_field_infty},
\end{equation}
allowing the optical extinction cross section to be written as
\begin{equation}
    \begin{aligned}
        \sigma_{\textrm{ext}}(\omega) = 4 \pi \frac{\omega}{c} \textrm{Im} \bigg\{ \int d \mathbf{x} \, d \mathbf{x}' \, \hat{\bf{J}}^{*}\cdot \tensor{\bf G}(\mathbf{x}, \mathbf{x}', \omega) \cdot \hat{\bf{J}}\bigg\}, \\
    \end{aligned} 
\end{equation}
where
\begin{equation}
    \tensor{\bf G}(\mathbf{x}, \mathbf{x}', \omega) = \delta(\mathbf{x} - \mathbf{x}_{\infty}) \frac{|\mathbf{x}_t - \mathbf{x}_{\infty}|}{|\mathbf{x}_t - \mathbf{x}|} e^{-ik|\mathbf{x}_t - \mathbf{x}|} \tensor{\bf I}\cdot \tensor{\boldsymbol{\alpha}}(\omega) \cdot \tensor{\bf I} \frac{|\mathbf{x}_t - \mathbf{x}_{\infty}|}{|\mathbf{x}_t - \mathbf{x}'|} e^{ik|\mathbf{x}_t - \mathbf{x}'|} \delta(\mathbf{x}' - \mathbf{x}_{\infty})
\end{equation}
and $\hat{\bf J}=\hat{\bm\epsilon}.$

Target responses beyond the single dipole approximation can be evaluated using the discrete dipole approximation \cite{draine2008discrete}, whereby the target is approximated as a collection of dipoles interacting self-consistently under a driving field $\mathbf{E}$. In this case, $\tensor{\bf G}(\vec{x}, \vec{x}', \omega)$ can be expanded as \cite{asenjo2014dichroism, lourencco2021optical}
\begin{equation}
    \tensor{\bf G}(\vec{x},\vec{x}', \omega) = \frac{-1}{4\pi\omega^2}\sum_{jj'}\tensor{\bf G}_0(\vec{x}-\vec{x}_j, \omega) \cdot  \bigg( \tensor{\boldsymbol{\alpha}}^{-1} - \tensor{\bf G}_{0} \bigg)_{jj'}^{-1} \cdot \tensor{\bf G}_0(\vec{x}_j'-\vec{x}', \omega),
    \label{G_exp_identity}
\end{equation}
or, in continuous form, as
\begin{equation}
    \tensor{\bf G}(\vec{x},\vec{x}', \omega) = \frac{-1}{4\pi\omega^2}\iint d\mathbf{y} d\mathbf{y}' \tensor{\bf G}_0(\vec{x},\vec{y}, \omega) \cdot \tensor{\boldsymbol{\alpha}}(\mathbf{y}, \mathbf{y}', \omega) \cdot \tensor{\bf G}_0(\vec{y}',\vec{x}', \omega).
    \label{G_exp_identity_cont}
\end{equation}

\section{Optical Polarization Analogs}\label{section_OPAs}
\subsection{General Considerations}\label{ssection_OPAs_general}

Suppose that the free electron wave function can be separated within an orthogonal coordinate system with variables $x_1, x_2, x_3$, i.e., $\psi(\vec{x}) = \chi(x_1)\zeta(x_2)\Phi(x_3)$, then the transverse transition current can be expressed as 
\begin{equation}
    \begin{aligned}
        \vec{J}_{fi}^{\perp}(\vec{x}) &= \frac{i \hbar e}{2 m} \bigg\{ \Psi^*_f(\vec{x}_{\perp}) \big[ \nabla_\perp \Psi_i(\vec{x}_\perp)  \big] - \Psi_i(\vec{x}_{\perp}) \big[ \nabla_\perp \Psi^*_f(\vec{x}_{\perp})  \big] \bigg\} \Psi_{f}^{*}(x_3)\Psi_i(x_3) \\
        &= \frac{i \hbar e}{2 m} \bigg\{ \chi^*_f(x_1)\zeta^*_f(x_2) \big[ \nabla_\perp \chi_i(x_1)\zeta_i(x_2)  \big] - \chi_i(x_1)\zeta_i(x_2) \big[ \nabla_\perp \chi^*_f(x_1)\zeta^*_f(x_2)  \big] \bigg\} \Phi_{f}^{*}(x_3)\Phi_i(x_3) \\
        &= \frac{i \hbar e}{2 m} \bigg\{ \chi^*_f(x_1)\zeta^*_f(x_2) \bigg[ \frac{1}{h_1} \frac{\partial \chi_i(x_1)}{\partial x_1} \zeta_i(x_2) \hat{\vec{x}}_1 + \frac{1}{h_2} \chi_i(x_1) \frac{\partial \zeta_i(x_2)}{\partial x_2} \hat{\vec{x}}_2  \bigg]  + \\
        &~~~~~~~~~~~~~~~~~~ -\chi_i(x_1)\zeta_i(x_2) \bigg[ \frac{1}{h_1} \frac{\partial \chi^*_f(x_1)}{\partial x_1} \zeta^*_f(x_2) \hat{\vec{x}}_1 + \frac{1}{h_2} \chi^*_f(x_1) \frac{\partial \zeta_f^*(x_2)}{\partial x_2} \hat{\vec{x}}_2  \bigg] \bigg\} \Phi_{f}^{*}(x_3)\Phi_i(x_3) \\
        &= \frac{i \hbar e}{2 m} \bigg\{ \frac{1}{h_1} \zeta^*_f(x_2)\zeta_i(x_2) \bigg[ \chi^*_f(x_1) \frac{\partial \chi_i(x_1)}{\partial x_1}   - \chi_i(x_1)  \frac{\partial \chi^*_f(x_1)}{\partial x_1} \bigg] \hat{\vec{x}}_1  + \\
        &~~~~~~~~~~~~~~~~~~ +  \frac{1}{h_2} \chi_f^*(x_1)\chi_i(x_1)\bigg[ \zeta_f^*(x_2) \frac{\partial \zeta_i(x_2)}{\partial x_2}  -  \zeta_i(x_2) \frac{\partial \zeta_f^*(x_2)}{\partial x_2}   \bigg]\hat{\vec{x}}_2 \bigg\} \Phi_{f}^{*}(x_3)\Phi_i(x_3) \\
    \end{aligned}
\end{equation}
where $h_i$ is the scale factor associated with coordinate $x_i$. It is evident that the $x_1$ component of the transition current density vanishes, for example, when 
\begin{equation}
    \chi^*_f(x_1) \frac{\partial \chi_i(x_1)}{\partial x_1}   - \chi_i(x_1)  \frac{\partial \chi^*_f(x_1)}{\partial x_1} =0.
\end{equation}
By writing the wave function in polar form $\chi(x_1) = A(x_1)e^{i\phi(x_1)}$ with $A(x_1) \in \mathbb{R}_{>0}$ and $\phi(x_1) \in [0,2\pi]$, this condition can be rewritten as 
\begin{equation}
     A_fe^{-i\phi_f} \bigg[ \frac{\partial A_i}{\partial x_1} + i\phi_i A_i \frac{\partial \phi_i}{\partial x_1} \bigg] e^{i \phi_i}   = A_ie^{i\phi_i} \bigg[ \frac{\partial A_f}{\partial x_1} - i\phi_f A_f \frac{\partial \phi_f}{\partial x_1} \bigg] e^{-i\phi_f}.
\end{equation}
If $f=i$ and $\phi_i = \phi_f = \phi$, then this becomes 
\begin{equation}
     \frac{\partial \phi}{\partial x_1}    = - \frac{\partial \phi}{\partial x_1}.
\end{equation}
This condition is satisfied if the phase $\phi$ is strictly constant. In summary, using the notation presented in the main text, the component of the transition current density along $\hat{\bf x}_j$ ($j=1,2$) vanishes provided (1) $\Psi_j(x_j)$ remains unchanged during the interaction, and (2) $\textrm{Arg}\big\{ \Psi_j(x_j) \big\}$ is constant.

\subsection{Linear Momentum OPA and Associated Transition Field $\vec{E}_{fi}^{0}(\vec{x}, \omega)$}\label{ssection_OPAs_LM}

When the initial and final electron states are plane waves with $ \langle \vec{x} | \vec{k}_j\rangle =  \psi_{j}(\vec{x}) = L^{-3/2} e^{i \vec{k}_j \cdot \vec{x}}$, the transition current density given by Eq. \eqref{J_fi} becomes (in units of statC$\cdot$s$^{-1}\cdot$cm$^{-2}$)
\begin{equation}
     \vec{J}_{fi}(\vec{x}) = -\frac{\hbar e}{2mL^3} (2\vec{k}_i - \vec{q}) e^{i \vec{q}\cdot \vec{x}}.
     \label{J_fi_pw}
\end{equation} 
Using the free space Green dyadic given by Eq. \eqref{G0_dyadic_def} in section \ref{ssection_FSOPW_cross_section}, the electric field sourced by the transition current density can be obtained. The induced electric field (in units of statV$\cdot$cm$^{-1}$ $\cdot$s) is
\begin{equation}
    \begin{aligned}
        \vec{E}_{fi}^{0}(\vec{x},\omega) & = -4 i \pi \omega\int d \vec{x}' \, \tensor{\bf G}_0(\mathbf{x}, \mathbf{x}', \omega) \cdot \bigg(\frac{L}{v} \bigg)\vec{J}_{fi}(\vec{x}').
    \end{aligned}
    \label{E_tr_freespace}
\end{equation}
Using the Fourier transform of the scalar Green's function $\int d \vec{x}'\,  \frac{e^{i\frac{\omega}{c}|\vec{x}' - \vec{x}|}}{|\vec{x}' - \vec{x}|} e^{\mp i \vec{q}\cdot \vec{x}'} = -\frac{4\pi}{\frac{\omega^2}{c^2} + q^2} e^{\mp i \vec{q}\cdot \vec{x}}$, we can perform the necessary spatial integration in Eq. \eqref{E_tr_freespace} to obtain an analytical form for the induced electric field (with units statV$\cdot$cm$^{-1}$$\cdot$s),
\begin{equation}
    \begin{aligned}
        \vec{E}_{fi}^{0}(\vec{x}, \omega ) & = \frac{i}{\omega} \int d \vec{x}' \, \Big\{\Big(\frac{\omega}{c}\Big)^2\tensor{\bf I} + \nabla \nabla \Big\}\frac{e^{i\frac{\omega}{c}|\vec{x} - \vec{x'}|}}{|\vec{x} - \vec{x'}|}  \cdot \bigg(\frac{L}{v} \bigg) \vec{J}_{fi}(\vec{x}') \\
        & = -\frac{i\hbar e}{2m v L^2 \omega} \Big\{\Big(\frac{\omega}{c}\Big)^2\tensor{\bf I} + \nabla \nabla \Big\} \cdot \Big\{ 
        2\vec{k}_i - \vec{q} \Big\} \int d \vec{x}' \, \frac{e^{i\frac{\omega}{c}|\vec{x} - \vec{x'}|}}{|\vec{x} - \vec{x'}|} e^{i \vec{q}\cdot \vec{x}'} \\
        & = \frac{2i \pi\hbar e}{m v L^2 \omega}  \frac{e^{i \vec{q}\cdot \vec{x}}}{\big(\frac{\omega}{c}\big)^2 + q^2} \Big\{\Big(\frac{\omega}{c}\Big)^2\tensor{\bf{I}} - \vec{q}\vec{q} \Big\} \cdot \Big\{ 2\vec{k}_i - \vec{q} \Big\} \\
        & = \frac{2i \pi  e \gamma_i}{k_i L^2 \omega}  \frac{e^{i \vec{q}\cdot \vec{x}}}{\big(\frac{\omega}{c}\big)^2 + q^2} \Big\{\Big(\frac{\omega}{c}\Big)^2\tensor{\bf{I}} - \vec{q}\vec{q} \Big\} \cdot \Big\{ 2\vec{k}_i - \vec{q} \Big\}. \\
    \end{aligned}
\end{equation}

\subsection{Orbital Angular Momentum OPA and Associated Transition Field $\vec{E}_{{fi}}^{0}(\vec{x}, \omega)$} \label{ssection_OPAs_OAM}

We can now turn our attention to acquiring the effective transition current density and the resulting electric field it sources when the electron undergoes transitions between Laguerre-Gauss and Hermite-Gauss transverse states. In the weak-focusing limit, these modes take the form \cite{allen1992orbital, lourencco2021optical, bourgeois2022polarization},
\begin{equation}
    \begin{aligned}
        \Psi_{n,m}^{\textrm{HG}}(x,y) & = \frac{1}{w_0}\sqrt{\frac{2}{\pi n! m!}}  \, 2^{-\frac{n+m}{2}} H_{n} \bigg[ \frac{x\sqrt{2}}{w_{0}} \bigg] H_{m} \bigg[ \frac{y\sqrt{2}}{w_{0}} \bigg] e^{-\frac{x^2 + y^2}{w_0^{2}}} \\
        \Psi^{\textrm{LG}}_{l,p}(R, \phi) & = \frac{1}{w_0} \sqrt{ \frac{2p!}{\pi( |\ell| + p)!} } \bigg( \frac{\sqrt{2}R}{w_0} \bigg)^{|\ell|} \textrm{L}_{p}^{(|\ell|)} \bigg[ \frac{2R^2}{w^{2}_0} \bigg] \, e^{i\ell \phi }e^{-\frac{x^2 + y^2}{w_0^{2}}}.
    \end{aligned}
    \label{LG_HG_wfl}
\end{equation}
Due to the following recurrence relations involving the Hermite-Gauss functions, $H_{n+1}\big[ \frac{x\sqrt{2}}{w_{0}} \big] =  \frac{2x\sqrt{2}}{w_0}H_{n} \big[ \frac{x\sqrt{2}}{w_{0}} \big] - \frac{w_0}{\sqrt{2}}H_{n}' \big[ \frac{x\sqrt{2}}{w_{0}} \big]$ and $
\frac{d}{dx} H_{n} \big[ \frac{x\sqrt{2}}{w_{0}} \big] = \frac{2n\sqrt{2}}{w_0}H_{n-1}\big[ \frac{x\sqrt{2}}{w_{0}} \big]$
one can derive a general expression for the derivative of an arbitrary Hermite-Gauss state $\partial_x \Psi_{n,m}^{\textrm{HG}} = \frac{1}{w_0}\big( \sqrt{n}\Psi_{n-1,m}^{\textrm{HG}} - \sqrt{n+1}\Psi_{n+1,m}^{\textrm{HG}} \big) $ and $\partial_y \Psi_{n,m}^{\textrm{HG}} = \frac{1}{w_0}\big( \sqrt{m}\Psi_{n,m-1}^{\textrm{HG}} - \sqrt{m+1}\Psi_{n,m+1}^{\textrm{HG}} \big)$.
With this in mind, we can construct an effective transition current density for the transition between any two Hermite-Gauss transverse electron states (units of statC$\cdot$s$^{-1}$ $\cdot$cm$^{-2}$)
\begin{equation}
    \begin{aligned}
        \vec{J}_{n,m \rightarrow n',m'}^{\textrm{HG}}(\vec{x}) & = -\frac{i \hbar e }{2mw_0 L} \bigg( \big\{ \sqrt{n}\Psi_{n-1,m}\Psi_{n',m'} - \sqrt{n+1}\Psi_{n+1,m}\Psi_{n',m'}  - \sqrt{n'}\Psi_{n,m}\Psi_{n'-1,m'} 
        + \sqrt{n'+1}\Psi_{n,m}\Psi_{n'+1,m'} \big\}\hat{\vec{x}} \\
        & + \big\{ \sqrt{m}\Psi_{n,m-1}\Psi_{n',m'}-\sqrt{m+1}\Psi_{n,m+1}\Psi_{n',m'}  -\sqrt{m'}\Psi_{n,m}\Psi_{n',m'-1} +\sqrt{m'+1}\Psi_{n,m}\Psi_{n',m'+1} \big\}\hat{\vec{y}} \\
        & + i\big\{ k_i^{z} + k_f^{z} \big\}w_0\Psi_{n,m}\Psi_{n',m'} \hat{\vec{z}} \bigg) e^{iq_z z}. \\
    \end{aligned}
    \label{J_HG}
\end{equation}
Since the first order Hermite and Laguerre-Gauss states are related via $\Psi_{0, \pm 1}^{\textrm{LG}} = 1/\sqrt{2} \big\{ \Psi_{10}^{\textrm{HG}} \pm i \Psi_{01}^{\textrm{HG}} \big\}$, we can construct a form for the effective transition current density  involving first order Laguerre-Gauss states $\mathbf{J}_{0, \pm 1 \rightarrow 00}^{\textrm{LG}}(\vec{x})$ in terms of $\mathbf{J}_{10 \rightarrow 00}^{\textrm{HG}}$ and $\mathbf{J}_{01 \rightarrow 00}^{\textrm{HG}}$ as
\begin{equation}
    \mathbf{J}_{0, \pm 1 \rightarrow 00}^{\textrm{LG}}(\vec{x}) = \frac{1}{\sqrt{2}} \Big\{ \mathbf{J}_{10 \rightarrow 00}^{\textrm{HG}} (\mathbf{x}) \pm i \mathbf{J}_{01 \rightarrow 00}^{\textrm{HG}} (\mathbf{x}) \Big\}.
    \label{J_LG}
\end{equation}
Using these equations, we explicitly write out the effective transition current density, Eq. \eqref{J_fi} for the various first order Hermite-Gauss and Laguerre-Gauss beams discussed in the main text
\begin{equation}
    \begin{aligned}
        \mathbf{J}_{00 \rightarrow 00}^{\textrm{HG}} (\mathbf{x})  & = \frac{\hbar e}{ 2 m L} \big|\Psi_{00}^{\textrm{HG}}\big|^2 (k_i + k_f)\hat{\vec{z}} e^{i q_z z} \\
        \mathbf{J}_{10 \rightarrow 00}^{\textrm{HG}} (\mathbf{x})  & = -\frac{i \hbar e}{ m w_0 L} \big|\Psi_{00}^{\textrm{HG}}\big|^2 \Big\{ \hat{\vec{x}} + i x(k_i + k_f)\hat{\vec{z}} \Big\}e^{i q_z z} \\
        \mathbf{J}_{01 \rightarrow 00}^{\textrm{HG}} (\mathbf{x}) & = -\frac{i \hbar e}{ m w_0 L} \big|\Psi_{00}^{\textrm{HG}}\big|^2 \Big\{  \hat{\vec{y}} + iy(k_i + k_f)\hat{\vec{z}} \Big\}e^{i q_z z} \\
        \mathbf{J}_{0, \pm 1 \rightarrow 00}^{\textrm{LG}}(\vec{x}) & = -\frac{i \hbar e}{m w_0 L} \big|\Psi_{00}^{\textrm{HG}}\big|^2 \, \Big\{ \hat{\vec{x}} \pm i \hat{\vec{y}} + i(k_i + k_f)\big( x \pm i y \big)\hat{\vec{z}} \Big\} e^{i q_z z}. 
    \end{aligned}
    \label{J_HGandLG}
\end{equation}
To find analytic solutions for the fields of the OAM states, we need to know the forms for the Hermite-Gauss modes and the product of two Hermite-Gauss modes in the limit that $w_0 \rightarrow 0$. Following previous work \cite{bourgeois2022polarization, lourencco2021optical}, and using the relation $\delta^{(n)}(x-x_0)\delta^{(n')}(x-x_0) = \frac{1}{w_0\sqrt{\pi}}\delta^{(n+n')}(x-x_0)$, one finds
\begin{equation}
    \begin{aligned}
        \lim_{w_0 \rightarrow 0^+} \Psi_{n,m}^{\textrm{HG}} & = \frac{w_0\sqrt{\pi}}{n! m!} \bigg( \frac{-w_0}{2} \bigg)^{(n+m)}\delta^{(n)}(x-x_0)\delta^{(m)}(y-y_0) \\
        \lim_{w_0 \rightarrow 0^+} (\Psi_{n',m'}^{\textrm{HG}})^*\Psi_{n,m}^{\textrm{HG}} & = \frac{1}{\sqrt{n!m!n'!m'!}} \bigg( \frac{-w_0}{2} \bigg)^{(n+m+n'+m')} \delta^{(n+n')}(x-x_0)\delta^{(m+m')}(y-y_0). \\
    \end{aligned}
    \label{limiting_HG}
\end{equation}
Using these limiting forms for the OAM state wavefunctions, the effective transition current densities take the form
\begin{equation}
    \begin{aligned}
        \mathbf{J}_{00 \rightarrow 00}^{\textrm{HG}} (\mathbf{x})  & = \frac{\hbar e}{ 2 m L} \delta^{(0)}(x-x_0)\delta^{(0)}(y-y_0)(k_i + k_f)\hat{\vec{z}} e^{i q_z z} \\
        \mathbf{J}_{10 \rightarrow 00}^{\textrm{HG}} (\mathbf{x})  & = -\frac{i \hbar e}{ m w_0 L} \Big\{ \delta^{(0)}(x-x_0)\delta^{(0)}(y-y_0)\hat{\vec{x}} - \frac{iw_0^2}{4}\delta^{(1)}(x-x_0)\delta^{(0)}(y-y_0)(k_i + k_f)\hat{\vec{z}} \Big\}e^{i q_z z} \\
        \mathbf{J}_{01 \rightarrow 00}^{\textrm{HG}} (\mathbf{x}) & = -\frac{i \hbar e}{ m w_0 L} \Big\{ \delta^{(0)}(x-x_0)\delta^{(0)}(y-y_0)\hat{\vec{y}} - \frac{iw_0^2}{4}\delta^{(0)}(x-x_0)\delta^{(1)}(y-y_0)(k_i + k_f)\hat{\vec{z}} \Big\}e^{i q_z z}  \\
        \mathbf{J}_{0, \pm 1 \rightarrow 00}^{\textrm{LG}}(\vec{x}) & = -\frac{i \hbar e}{m w_0 L} \, \Big\{ \delta^{(0)}(x-x_0)\delta^{(0)}(y-y_0) (\hat{\vec{x}} \pm i \hat{\vec{y}}) \\
        & \quad \quad \quad \quad \quad \quad - \frac{iw_0^2}{4}(k_i + k_f)\big( \delta^{(1)}(x-x_0)\delta^{(0)}(y-y_0) \pm i \delta^{(0)}(x-x_0)\delta^{(1)}(y-y_0) \big)\hat{\vec{z}} \Big\} e^{i q_z z}. 
    \end{aligned}
    \label{J_HGandLG_w0tozero}
\end{equation}
Making use of the identity $\int dz' \, \frac{e^{i(\frac{\omega}{c})|\vec{x} - \vec{x}'|}}{|\vec{x} - \vec{x}'|}e^{\pm iq_z z'} = 2\textrm{K}_{0}(\frac{q_z \Delta \vec{R}}{\gamma})e^{\pm i q_z z}$, we can use the OAM transition current densities to find effective transition electromagnetic fields. For instance, the electric field for the $\Psi_{1, 0}^{\textrm{HG}} \rightarrow \Psi_{0,0}^{\textrm{HG}}$ transition is 
\begin{equation}
    \begin{aligned}
        \vec{E}_{10 \rightarrow 00}^{0}(\vec{x}, \omega ) & = \, -4 \pi i \omega \int d\mathbf{x}' \, \tensor{\mathbf{G}}_{0}(\vec{x}, \vec{x}', \omega) \cdot  \Big( \frac{L}{v} \Big) \, \mathbf{J}_{10 \rightarrow 00}^{\textrm{HG}} (\mathbf{x}') \\
        & = \frac{i}{\omega} \int d \vec{x}' \, \Big\{\Big(\frac{\omega}{c}\Big)^2\tensor{\bf I} + \nabla \nabla \Big\}\frac{e^{i\frac{\omega}{c}|\vec{x} - \vec{x'}|}}{|\vec{x} - \vec{x'}|}  \cdot \Big( \frac{L}{v} \Big) \, \mathbf{J}_{10 \rightarrow 00}^{\textrm{HG}} (\mathbf{x}') \\
        & = \frac{i}{\omega}\frac{-i\hbar e}{ m w_0 v} \Big\{\Big(\frac{\omega}{c}\Big)^2\tensor{\bf I} + \nabla \nabla \Big\} \cdot \int d \vec{x}' \,\frac{e^{i\frac{\omega}{c}|\vec{x} - \vec{x'}|} e^{i q_z z'}}{|\vec{x} - \vec{x'}|} \Big\{ \delta^{(0)}(x'-x_0)\delta^{(0)}(y'-y_0)\hat{\vec{x}} \\
        & \quad \quad \quad \quad \quad \quad \quad \quad \quad \quad \quad \quad \quad \quad \quad \quad \quad \quad \quad \quad \quad \quad \quad \quad  -\frac{iw_0^2}{4}\delta^{(1)}(x'-x_0)\delta^{(0)}(y'-y_0)(k_i + k_f)\hat{\vec{z}} \Big\} \\
        & = \frac{ic}{\omega}\Big\{\Big(\frac{\omega}{c}\Big)^2\tensor{\bf I} + \nabla \nabla \Big\} \cdot \Big( \frac{-2 i \hbar e}{m w_0 vc}\Big)\Big\{ \textrm{K}_{0}\Big(\frac{q_z \Delta \vec{R}_{0} }{\gamma}\Big) \hat{\vec{x}} + \frac{iq w_0^2}{4\gamma}(\hat{\Delta \vec{R}_{0}} 
        \cdot \hat{\vec{x}})(k_i + k_f)\textrm{K}_{1}\Big(\frac{q_z \Delta \vec{R}_{0}}{\gamma}\Big)\hat{\vec{z}} \Big\} e^{i q_z z} \\
        & = \frac{ic}{\omega}\Big\{\Big(\frac{\omega}{c}\Big)^2\tensor{\bf I} + \nabla \nabla \Big\} \cdot \vec{A}_{10 \rightarrow 00}^{0}(\vec{x}, \omega ).
    \end{aligned}
\end{equation}
where we have defined an effective transition vector potential $\vec{A}_{10 \rightarrow 00}^{0}(\vec{x}, \omega ) = -\Big( \frac{2 i \hbar e}{m w_0 vc}\Big)\Big\{ \textrm{K}_{0}\Big(\frac{q_z \Delta \vec{R}_{0} }{\gamma}\Big) \hat{\vec{x}} + \frac{iq_z w_0^2}{4\gamma}(\hat{\Delta \vec{R}_{0}} \cdot \hat{\vec{x}})(k_i + k_f)\textrm{K}_{1}\Big(\frac{q_z \Delta \vec{R}_{0}}{\gamma}\Big)\hat{\vec{z}} \Big\} e^{i q_z z}$ in units of StatV$\cdot$s and where the impact parameter is $\Delta \vec{R}_{0} = |\vec{R} - \vec{R}_0|$. Repeating this process for the $\mathbf{J}_{00 \rightarrow 00}^{\textrm{HG}}$ and $\mathbf{J}_{01 \rightarrow 00}^{\textrm{HG}}$ electron state transitions, we find that (statV$\cdot$cm$^{-1}$$\cdot$s)
\begin{equation}
    \begin{aligned}
        \vec{E}_{00 \rightarrow 00}^{0}(\vec{x}, \omega ) & = \frac{ic}{\omega}\Big\{\Big(\frac{\omega}{c}\Big)^2\tensor{\bf I} + \nabla \nabla \Big\} \cdot \Big( \frac{ \hbar e}{m vc}\Big)(k_i + k_f)\textrm{K}_{0}\Big(\frac{q_z \Delta \vec{R}_{0}}{\gamma}\Big)\hat{\vec{z}} \,  e^{i q_z z} \\
        \vec{E}_{10 \rightarrow 00}^{0}(\vec{x}, \omega ) & = \frac{ic}{\omega}\Big\{\Big(\frac{\omega}{c}\Big)^2\tensor{\bf I} + \nabla \nabla \Big\} \cdot \Big( \frac{-2 i \hbar e}{m w_0 vc}\Big)\Big\{ \textrm{K}_{0}\Big(\frac{q_z \Delta \vec{R}_{0} }{\gamma}\Big) \hat{\vec{x}} + \frac{iq_z w_0^2}{4\gamma}(\hat{\Delta \vec{R}_{0}} 
        \cdot \hat{\vec{x}})(k_i + k_f)\textrm{K}_{1}\Big(\frac{q_z \Delta \vec{R}_{0}}{\gamma}\Big)\hat{\vec{z}} \Big\} e^{i q_z z} \\
        \vec{E}_{01 \rightarrow 00}^{0}(\vec{x}, \omega ) & = \frac{ic}{\omega}\Big\{\Big(\frac{\omega}{c}\Big)^2\tensor{\bf I} + \nabla \nabla \Big\} \cdot \Big( \frac{-2 i \hbar e}{m w_0 vc}\Big)\Big\{ \textrm{K}_{0}\Big(\frac{q_z \Delta \vec{R}_{0} }{\gamma}\Big) \hat{\vec{y}} + \frac{iq_z w_0^2}{4\gamma}(\hat{\Delta \vec{R}_{0}} 
        \cdot \hat{\vec{y}})(k_i + k_f)\textrm{K}_{1}\Big(\frac{q_z \Delta \vec{R}_{0}}{\gamma}\Big)\hat{\vec{z}} \Big\} e^{i q_z z}.
    \end{aligned}
    \label{E_HGandLG_w0tozero}
\end{equation}
Due to the relation between the first order HG- and LG- states, (see Eq. \eqref{J_LG} earlier in this section) linear combinations of $\vec{E}_{10 \rightarrow 00}^{0}(\vec{x}, \omega )$ and $\vec{E}_{01 \rightarrow 00}^{0}(\vec{x}, \omega )$ will produce the transition fields due to transitions between first order LG states $\vec{E}_{0,\pm 1 \rightarrow 00}^{0}(\vec{x}, \omega )$.

\section{Fully-Retarded Double Differential Inelastic Scattering Cross Section}\label{section_DDSCS}

In this section, we will derive the double differential scattering cross section for electron plane waves, Eq. (2) in the manuscript. The total relativistic energy (kinetic energy + rest mass energy) of the incoming free electron is $E_i = \gamma_i m c^2$, where $\gamma_i = (1 - \beta_i^2)^{-1/2}$ is the Lorentz contraction factor, and $\beta = v/c$. The form of the final electron energy is identical with $i\to f$, and $E_{if} = E_i - E_f = (\gamma_i - \gamma_f) mc^2$. Knowing the loss energy $E_{if}$ and the incoming electron speed, the final electron speed can be determined using $\gamma_f = \gamma_i - (E_{if}/mc^2)$ and $\beta_f = \sqrt{1 - \gamma_f^{-2}}$. This fixes the magnitude of the outgoing wave vector $k_f = m \gamma_f \beta_f c / \hbar $. The observation angle $\theta$ is then selected such that $\mathbf{k}_f = k_f \cos \theta ~\hat{\mathbf{z}} + k_f \sin \theta ~\hat{\mathbf{x}}_{\perp}$, and $\mathbf{q} = (k_i - k_f \cos \theta) \hat{\mathbf{z}} -  k_f \sin \theta ~\hat{\mathbf{x}}_{\perp} $.

Inserting the transition current density for plane wave electron states given by Eq. \eqref{J_fi_pw} into Eq. \eqref{w_fi_loss_res_altform}, defining $\tilde{\vec{k}} = 2 \vec{k}_i - \vec{q} = \vec{k}_i + \vec{k}_f$, and summing over final electron states with $\sum_{\vec{k}_{f}} \rightarrow \big( L/2\pi\big)^3 \int d\vec{k}_f$, one finds (which is unitless)
\begin{equation}
    \begin{aligned}
        w^{\textrm{loss}} (\omega ) & = -\frac{8 \pi}{\hbar} \sum_{\vec{k}_{f}} \int d\vec{x} \, d\vec{x}' \, \textrm{Im}\bigg\{ \vec{J}_{fi}^*(\vec{x}) \cdot \tensor{\mathbf{G}}(\vec{x}, \vec{x}', \omega)  \cdot\vec{J}_{fi}(\vec{x}') \bigg\}\delta (\omega - \omega_{if}) \\ 
        & = -\frac{8 \pi}{\hbar} \bigg( \frac{\hbar e}{2mL^3} \bigg)^2 \bigg( \frac{L}{2\pi} \bigg)^3 \int d\vec{k}_f \, d\vec{x} \, d\vec{x}' \, \textrm{Im}\bigg\{ e^{-i \vec{q}\cdot \vec{x}} \tilde{\vec{k}}^* \cdot \tensor{\mathbf{G}}(\vec{x}, \vec{x}', \omega)  \cdot \tilde{\vec{k}} e^{i \vec{q}\cdot \vec{x}'} \bigg\}\delta (\omega - \omega_{if}) \\
        & = -\frac{\hbar e^2}{4 \pi^2 m^2 L^3} \int d\vec{k}_f \, d\vec{x} \, d\vec{x}' \, \textrm{Im}\bigg\{ e^{-i \vec{q}\cdot \vec{x}} \tilde{\vec{k}}^* \cdot \tensor{\mathbf{G}}(\vec{x}, \vec{x}', \omega)  \cdot \tilde{\vec{k}} e^{i \vec{q}\cdot \vec{x}'} \bigg\}\delta (\omega - \omega_{if}).
    \end{aligned}
\end{equation}
To arrive at the scattering cross section from the frequency-resolved transition rate directly above, we can divide $w^{\textrm{loss}} (\omega )$, by the incoming particle flux, $v/L^3$, yielding (with the expected units of cm$^{2}$$\cdot$s)
\begin{equation}
    \begin{aligned}
        \sigma^{\textrm{loss}} (\omega ) & = -\frac{8 \pi}{\hbar} \bigg( \frac{\hbar e}{2mL^3} \bigg)^2 \bigg( \frac{L}{2\pi} \bigg)^3 \frac{L^3}{v}\int d\vec{k}_f \, d\vec{x} \, d\vec{x}' \, \textrm{Im}\bigg\{ e^{-i \vec{q}\cdot \vec{x}} \tilde{\vec{k}}^* \cdot \tensor{\mathbf{G}}(\vec{x}, \vec{x}', \omega)  \cdot \tilde{\vec{k}} e^{i \vec{q}\cdot \vec{x}'} \bigg\}\delta (\omega - \omega_{if}) \\
        & = -\frac{\hbar e^2}{4 m^2 \pi^2 v} \int d\vec{k}_f \, d\vec{x} \, d\vec{x}' \, \textrm{Im}\bigg\{ e^{-i \vec{q}\cdot \vec{x}} \tilde{\vec{k}}^* \cdot \tensor{\mathbf{G}}(\vec{x}, \vec{x}', \omega)  \cdot \tilde{\vec{k}} e^{i \vec{q}\cdot \vec{x}'} \bigg\}\delta (\omega - \omega_{if}).
    \end{aligned}
    \label{sigma_loss_freqres}
\end{equation}

\subsection{Recovery of Well Known Form for an Isolated Dipolar Target in the Quasistatic Limit}\label{ssection_DDSCS_dipole}

For an isolated dipolar target with polarizability $\tensor{\boldsymbol{\alpha}}(\omega)$ located at position $\vec{x}_d$, the induced Green's function can be expressed as \cite{asenjo2014dichroism}
\begin{equation}
    \tensor{\mathbf{G}}(\vec{x}, \vec{x}', \omega) = -\frac{1}{4\pi \omega^2}\tensor{\mathbf{G}}^0(\vec{x}, \vec{x}_d, \omega) \cdot  \tensor{\boldsymbol{\alpha}}(\omega) \cdot \tensor{\mathbf{G}}^0(\vec{x}_d, \vec{x}', \omega), 
\end{equation}
and once inserted into Eq. \eqref{sigma_loss_freqres}, the scattering cross section for a dipole is 
\begin{equation}
    \begin{aligned}
        \sigma_{\textrm{dip}}^{\textrm{loss}} (\omega ) & = -\frac{\hbar e^2}{4 m^2 \pi^2 v} \int d\vec{k}_f \, d\vec{x} \, d\vec{x}' \, \textrm{Im}\bigg\{ e^{-i \vec{q}\cdot \vec{x}} \tilde{\vec{k}}^* \cdot \tensor{\mathbf{G}}(\vec{x}, \vec{x}', \omega)  \cdot \tilde{\vec{k}} e^{i \vec{q}\cdot \vec{x}'} \bigg\}\delta (\omega - \omega_{if}) \\
        & = \frac{\hbar e^2}{4 m^2 \pi^2 v} \int d\vec{k}_f \, d\vec{x} \, d\vec{x}' \, \frac{1}{4\pi \omega^2}\textrm{Im}\bigg\{ e^{-i \vec{q}\cdot \vec{x}} \tilde{\vec{k}}^* \cdot \tensor{\mathbf{G}}^0(\vec{x}, \vec{x}_d, \omega) \cdot  \tensor{\boldsymbol{\alpha}}(\omega) \cdot \tensor{\mathbf{G}}^0(\vec{x}_d, \vec{x}', \omega) \cdot \tilde{\vec{k}} e^{i \vec{q}\cdot \vec{x}'} \bigg\}\delta (\omega - \omega_{if}) \\
        & = \frac{\hbar e^2}{4 m^2 \pi^2 v} \int d\vec{k}_f \, \frac{1}{4\pi \omega^2}\textrm{Im}\bigg\{ \tilde{\vec{k}}^* \cdot \bigg[ \int d\vec{x} \,  e^{-i \vec{q} \vec{x}} \tensor{\mathbf{G}}^0(\vec{x}, \vec{x}_d, \omega)\bigg] \cdot  \tensor{\boldsymbol{\alpha}}(\omega) \cdot \bigg[ \tilde{\vec{k}} \cdot \int d \vec{x}'\tensor{\mathbf{G}}^0(\vec{x}_d, \vec{x}', \omega) e^{i \vec{q}\cdot \vec{x}'} \bigg] \bigg\}\delta (\omega - \omega_{if}).
    \end{aligned}
\end{equation}
Using the Fourier transform of the scalar Green's function defined in the previous section one can evaluate the following, $\int d\vec{x} \, e^{-i \vec{q}\cdot \vec{x}} \tensor{\mathbf{G}}^0(\vec{x}, \vec{x}_d, \omega) = -4\pi \bigg( \frac{\omega^2}{c^2}\tensor{\bf I} + \vec{q}\vec{q}\bigg) \frac{e^{-i \vec{q}\cdot \vec{x}_d}}{ \frac{\omega^2}{c^2} - q^2}$. Evaluating the integrals over $d\vec{x}$ and $d\vec{x}'$ we find (in units of cm$^2 \cdot$s) 
\begin{equation}
    \begin{aligned}
        \sigma_{\textrm{dip}}^{\textrm{loss}} (\omega ) & = \frac{\hbar e^2}{4 m^2 \pi^2 v} \int d\vec{k}_f \, \frac{16\pi^2}{4\pi \omega^2}\textrm{Im}\bigg\{ \tilde{\vec{k}}^* \cdot \bigg( \frac{\omega^2}{c^2}\tensor{\bf I} + \vec{q}\vec{q}\bigg)^*\frac{e^{-i \vec{q}\cdot \vec{x}_d}}{ \frac{\omega^2}{c^2} - q^2} \cdot  \tensor{\boldsymbol{\alpha}}(\omega) \cdot \tilde{\vec{k}} \cdot \bigg( \frac{\omega^2}{c^2}\tensor{\bf I} + \vec{q}\vec{q}\bigg)\frac{e^{i \vec{q}\cdot \vec{x}_d}}{ \frac{\omega^2}{c^2} - q^2} \bigg\}\delta (\omega - \omega_{if}) \\
        & = \frac{\hbar e^2}{m^2 \pi v} \int d\vec{k}_f \, \frac{1}{\omega^2}\textrm{Im}\bigg\{ \tilde{\vec{k}}^* \cdot \bigg( \frac{\omega^2}{c^2}\tensor{\bf I} + \vec{q}\vec{q}\bigg)^*\frac{1}{ \frac{\omega^2}{c^2} - q^2} \cdot  \tensor{\boldsymbol{\alpha}}(\omega) \cdot \tilde{\vec{k}} \cdot \bigg( \frac{\omega^2}{c^2}\tensor{\bf I} + \vec{q}\vec{q}\bigg)\frac{1}{ \frac{\omega^2}{c^2} - q^2} \bigg\}\delta (\omega - \omega_{if}).
    \end{aligned}
    \label{sigma_loss_freqres_dipole}
\end{equation}
In the quasistatic limit Eq. \eqref{sigma_loss_freqres_dipole} reduces to
\begin{equation}
    \begin{aligned}
        \lim_{c \rightarrow \infty}\sigma_{\textrm{dip}}^{\textrm{loss}} (\omega ) & = \frac{\hbar e^2}{m^2 \pi v} \int d\vec{k}_f \, \frac{1}{\omega^2 q^4}\textrm{Im}\bigg\{\big[ \tilde{\vec{k}}^* \cdot \vec{q}^* \big] \vec{q}^* \cdot  \tensor{\boldsymbol{\alpha}}(\omega) \cdot \big[ \tilde{\vec{k}} \cdot \vec{q} \big]\vec{q} \bigg\}\delta (\omega - \omega_{if}) \\
        & = \frac{\hbar e^2}{m^2 \pi v} \iiint d^2 \Omega \, dk_f \, k_f^2 \, \frac{1}{\omega^2 q^4} \big| \tilde{\vec{k}} \cdot \vec{q} \big|^2 \, \textrm{Im}\bigg\{\vec{q}^* \cdot  \tensor{\boldsymbol{\alpha}}(\omega) \cdot \vec{q} \bigg\}\delta (\omega - \omega_{if}).
    \end{aligned}
\end{equation}
Knowing that $| \tilde{\vec{k}} \cdot \vec{q} | = |\vec{k}_i^2 - \vec{k}_f^2| = \frac{2m\omega_{if}}{\hbar}$, we find for the scattering cross section,
\begin{equation}
    \begin{aligned}
        \frac{\partial \sigma_{\textrm{dip}}^{\textrm{loss}}}{\partial \Omega} & = \frac{\hbar e^2}{m^2 \pi v} \iint d\omega dk_f \, k_f^2 \, \frac{1}{\omega^2 q^4} \big| \tilde{\vec{k}} \cdot \vec{q} \big|^2 \, \textrm{Im}\bigg\{\vec{q}^* \cdot  \tensor{\boldsymbol{\alpha}}(\omega) \cdot \vec{q} \bigg\}\delta (\omega - \omega_{if}) \\
        & = \frac{4e^2}{\hbar \pi v} \iint d\omega dk_f \, k_f^2 \, \frac{\omega_{if}^2}{\omega^2 q^4} \, \textrm{Im}\bigg\{\vec{q}^* \cdot  \tensor{\boldsymbol{\alpha}}(\omega) \cdot \vec{q} \bigg\}\delta (\omega - \omega_{if}) \\
        & = \frac{4e^2}{\hbar \pi v} \int  dk_f \, k_f^2 \, \frac{1}{ q^4} \, \textrm{Im}\bigg\{\vec{q}^* \cdot  \tensor{\boldsymbol{\alpha}}(\omega_{if}) \cdot \vec{q} \bigg\}.
    \end{aligned}
\end{equation}
Changing variables from momentum to energy using $dE_{if} = -(\frac{\hbar^2 k_f}{m})dk_f$, we obtain for the double differential scattering cross section in the quasistatic limit
\begin{equation}
        \frac{\partial^2 \sigma}{\partial E_{if} \partial \Omega } = \frac{4 m^2 e^2}{q^4 \hbar^4 \pi} \bigg( \frac{k_f}{k_i} \bigg) \, \textrm{Im}\Big\{\vec{q}^* \cdot  \tensor{\boldsymbol{\alpha}}(E_{if} / \hbar) \cdot \vec{q}\bigg\}.
    \label{ddsigma_fi_dipole_quasistatic}
\end{equation}
We emphasize that this expression is derived by taking a macroscopic QED perspective of the target, with fully-retarded electromagnetic coupling between target and probing free electron. This is in contrast to the more conventional approach, where a microscopic perspective is taken with $j$ bound target electrons interacting with the probing free electron via the Coulomb interaction. The microscopic approach, in the small target $(\mathbf{q}\cdot \mathbf{x}_{j} \ll 1)$ limit, leads to the ``dipole approximation" for EELS \cite{Hebert2003-mn, Schattschneider2006-fy}, where the double differential cross section for electron scattering from a target with a single bound electron is  [cm$^2 \cdot$erg$^{-1}$]
\begin{equation}
    \begin{aligned}
        \frac{\partial^2 \sigma}{\partial E_{if} \partial \Omega } &= \sum_{b} \frac{4}{a_0^2} \bigg( \frac{k_f}{k_i}\bigg) \frac{1}{q^4} \big|\bra{b} i \vec{q} \cdot \hat{\vec{x}} \ket a \big|^2 \delta(E_b - E_a - E_{if}). \\
    \end{aligned}
\end{equation}
Here, $a$ is the target ground state, $b$ labels the target excited states. Noting that $\big|\bra{b} i \vec{q} \cdot \hat{\vec{x}} \ket{a} \big|^2 = \big( \bra{b} i \vec{q} \cdot \hat{\vec{x}} \ket{a} \big)^* \bra{b} i \vec{q} \cdot \hat{\vec{x}} \ket{a} = -i \vec{q}^* \cdot \bra{a}\hat{\vec{x}} \ket{b}  \bra{b}\hat{\vec{x}} \ket{a} \cdot  i \vec{q}   = q_l^* \bra{a}\hat{x}_{l} \ket{b}  \bra{b}\hat{x}_{j} \ket{a} q_{j}$, and using the relationship $ \frac{1}{\omega \pm i \eta} = \mathcal{P}\bigg(\frac{1}{\omega} \bigg) \mp i \pi \delta(\omega)$, it is possible to write
\begin{equation}
    \begin{aligned}
        \sum_b \big|\bra{b} i \vec{q} \cdot \hat{\vec{x}} \ket{a} \big|^2 \delta(E_{if} - E_{ba}) &= \frac{ 1 }{\pi e^2} ~q_l^*~ \textrm{Im} \bigg\{ \sum_b  \frac{ \bra{a}\hat{\mu}_{l} \ket{b}  \bra{b}\hat{\mu}_{j} \ket{a} }{ E_{if} - E_{ba} } \bigg\}~ q_{j}  \\
        &= \frac{ 1 }{\pi e^2} ~q_l^*~ \textrm{Im} \bigg\{ \alpha^{aa}_{lj}(\omega_{if}) \bigg\}~ q_{j}  \\
        &= \frac{ 1 }{\pi e^2} ~\vec{q}^* \cdot \textrm{Im} \bigg\{ \tensor{\boldsymbol{\alpha}}^{aa}(E_{if}) \bigg\} \cdot \vec{q} ,  \\
    \end{aligned}
\end{equation}
where $\hat{\mu}_{l} = e \hat{x}_l$ is the $l$th Cartesian component of the dipole operator and $\tensor{\boldsymbol{\alpha}}^{aa}(E_{if})$ is the ground state Kramers-Heisenberg polarizability tensor evaluated at the loss energy $E_{if}$. Putting this all together, we find the double differential inelastic scattering cross section for a single bound target electron  within the quasistatic dipole approximation to be [cm$^2\cdot$ erg$^{-1}$]
\begin{equation}
    \begin{aligned}
        \frac{\partial^2 \sigma^{\textrm{dip}}}{\partial E_{if} \partial \Omega } &= \frac{4 m^2 e^2}{\pi  \hbar^4} \bigg( \frac{k_f}{k_i}\bigg) \frac{1}{q^4} ~\vec{q}^* \cdot \textrm{Im} \bigg\{ \tensor{\boldsymbol{\alpha}}^{aa}(E_{if}) \bigg\} \cdot \vec{q} , \\
    \end{aligned}
    \label{diff_CS_quasistatic}
\end{equation}
which is in agreement with the quasistatic $c \to \infty$ limit form of $\partial^2 \sigma /\partial E_{if} \partial \Omega$ given by Eq. \eqref{ddsigma_fi_dipole_quasistatic}.

\section{Fully-Retarded State- and Energy-resolved EEL Probability $\Gamma_{fi}(\vec{R}, \omega)$}\label{section_gammaloss}

The energy-resolved loss probability is found by multiplying the frequency-resolved loss rate in Eq. \eqref{w_fi_loss_res_altform} by $L/\hbar v$ and summing over all the possible final electron momenta directed along the TEM axis (with $k_f^z \parallel q_{\parallel}$) and in units of erg$^{-1}$,
\begin{equation}
    \begin{aligned}
        \Gamma_{fi} (\vec{R}_0, \omega ) & = \Big( \frac{L^2}{2\pi \hbar v} \Big) \int d q_{\parallel} \, w^{\textrm{loss}}_{fi} (\omega ) \\ 
        & = \Big( \frac{2 \pi L^2}{\hbar^2 \omega v} \Big) \int d q_{\parallel} \, d\vec{x} \, d\vec{x}' \, \vec{J}_{fi}^*(\vec{x}) \cdot  \tensor{\boldsymbol{\varrho}}(\vec{x}, \vec{x}', \omega) \cdot\vec{J}_{fi}(\vec{x}') \delta ( \omega - \omega_{if}).
    \end{aligned}
    \label{gamma_fi_form1}
\end{equation}
The EEL probability can be re-expressed in terms of the induced electric field as show by Eq. (3) in the main text. This is done by using the property shown in section \ref{section_Derivation_MTEq1} for the Helmholtz Green's tensor and then using the following form for the induced electric field, ( $ \vec{E}_{fi}(\vec{x},\omega) = -4 i \pi \omega\int d \vec{x}' \, \tensor{\bf G}(\mathbf{x}, \mathbf{x}', \omega) \cdot (L/v) \vec{J}_{fi} \, (\vec{x}') $ ), to obtain 
\begin{equation}
    \begin{aligned}
        \Gamma_{fi} (\vec{R}_0, \omega )
        & = \Big( \frac{2 \pi L^2}{\hbar^2 \omega v} \Big) \int d q_{\parallel} \, d\vec{x} \, d\vec{x}' \, \vec{J}_{fi}^*(\vec{x}) \cdot  \bigg[ -\frac{2\omega}{\pi} \textrm{Im}\bigg\{ \tensor{\mathbf{G}}(\vec{x}, \vec{x}', \omega) \bigg\}\bigg] \cdot \vec{J}_{fi}(\vec{x}') \delta ( \omega - \omega_{if}) \\
        & = -\Big( \frac{4 v}{\hbar^2} \Big) \int d q_{\parallel} \, \textrm{Im}\bigg\{ \int d\vec{x} \, \Big( \frac{L}{v} \Big)\, \vec{J}_{fi}^*(\vec{x}) \cdot \int  d\vec{x}' \, \tensor{\mathbf{G}}(\vec{x}, \vec{x}', \omega) \Big( \frac{L}{v} \Big) \, \cdot\vec{J}_{fi}(\vec{x}') \bigg\}  \delta ( \omega - \omega_{if}) \\
        & = -\Big( \frac{4v}{\hbar^2 } \Big) \int d q_{\parallel} \, \textrm{Im}\bigg\{ -\frac{1}{4 i \pi \omega} \int d\vec{x} \, \Big( \frac{L}{v} \Big) \, \vec{J}_{fi}^*(\vec{x}) \cdot \vec{E}_{fi}(\vec{x},\omega) \bigg\}  \delta ( \omega - \omega_{if}) \\
        & = \int d q_{\parallel} \, \textrm{Re}\bigg\{ - \frac{v}{\pi \hbar^2 \omega} \int d\vec{x} \, \Big( \frac{L}{v} \Big) \, \vec{J}_{fi}^*(\vec{x}) \cdot \vec{E}_{fi}(\vec{x},\omega) \bigg\}  \delta ( \omega - \omega_{if}).
    \end{aligned}
    \label{Gamma_loss_fi_deltafxn}
\end{equation}

To arrive at Eq. (3) in the main text, we consider scattering interactions within the non-recoil approximation, where the momentum change of the STEM electron is primarily parallel to the electrons direction of propagation \cite{de2010optical}. Disregarding relativistic corrections, the energy change of the swift electron (assuming $v \parallel \vec{\hat{z}}$) is, $E_{if} = \frac{m v^2}{2} - \frac{| m \vec{v} - \hbar \vec{q}|^2}{2 m} = \hbar \vec{q} \cdot \vec{v} - \frac{\hbar^2 \vec{q}^2}{2 m} \approx \hbar (\vec{q} \cdot \vec{v})$. As a result, $\omega_{if}$ inside the energy conserving delta function in Eq. \eqref{Gamma_loss_fi_deltafxn} approximates to $\omega_{if} \approx \vec{q} \cdot v\hat{\vec{z}} = q_{\parallel} v$. Lastly, by changing variables from $\vec{x} \rightarrow (\vec{R}, z)$ and then integrating over $d q_{\parallel}$, we acquire the following form for the EEL probability (in units of erg$^{-1}$)
\begin{equation}
    \begin{aligned}
        \Gamma_{fi} (\vec{R}_0, \omega ) & = \int d q_{\parallel} \, \textrm{Re}\bigg\{ - \frac{v}{\pi \hbar^2 \omega}\int d\vec{R} \, dz  \, \Big( \frac{L}{v} \Big) \, \vec{J}_{fi}^*(\vec{R}, z) \cdot \vec{E}_{fi}(\vec{R}, z, \omega) \bigg\}  \delta ( \omega - q_{\parallel} v) \\
        & = \int d q_{\parallel} \, \textrm{Re}\bigg\{ - \frac{1}{\pi \hbar^2 \omega}\int d\vec{R} \, dz  \, \Big( \frac{L}{v} \Big) \, \vec{J}_{fi}^*(\vec{R}, z) \cdot \vec{E}_{fi}(\vec{R}, z, \omega) \bigg\}  \delta ( q_{\parallel} - \frac{\omega}{v}) \\
        & = \frac{1}{\pi \hbar^2 \omega} \textrm{Re}\bigg\{ - \int d\vec{R} \, dz  \, \Big( \frac{L}{v} \Big) \, \vec{J}_{fi}^*(\vec{R}, z) \cdot \vec{E}_{fi}(\vec{R}, z, \omega) \bigg\}. 
    \end{aligned}
    \label{Gamma_loss_fi}
\end{equation}

\subsection{$\Gamma_{fi}(\mathbf{R}, \omega)$ in the Narrow Beam Limit}\label{ssection_gammaloss_NBL}
When the beam waist of the electron probe $w_{0}$ is small compared to the length scale over which the response field changes, the electric field experienced by the target due to the presence of the fast electron will be approximately constant, i.e., $\vec{E}_{fi}(\vec{R}, z, \omega) \approx \vec{E}_{fi}(\vec{R}_0, z, \omega)$. This produces an EEL probability of the form $\Gamma_{fi} (\vec{R}_0, \omega ) \approx \frac{1}{\pi \hbar \omega} \textrm{Re}\big\{ -\int d\vec{R} \, dz  \, \Big( \frac{L}{v} \Big) \, \vec{J}_{fi}^*(\vec{R}, z) \cdot \vec{E}_{fi}(\vec{R}_0, z, \omega) \big\}$. Therefore, we arrive at the form for the EEL probability in the narrow beam width limit
\begin{equation}
    \begin{aligned}
        \Gamma_{fi} (\vec{R}_0, \omega ) & \approx \frac{1}{\pi \hbar^2 \omega} \textrm{Re}\bigg\{ - \int d\vec{R} \, dz  \, \Big( \frac{L}{v} \Big) \, \vec{J}_{fi}^*(\vec{R}, z) \cdot \vec{E}_{fi}(\vec{R}_0, z, \omega) \bigg\} \\
        & = \frac{1}{\pi \hbar^2 \omega} \textrm{Re}\bigg\{ - \Big( \frac{L}{v} \Big) \int dz  \, \Big[\int  d\vec{R} \, \vec{J}_{fi}^*(\vec{R}, z) \Big] \, \cdot \vec{E}_{fi}(\vec{R}_0, z, \omega) \bigg\} \\
        & = \frac{1}{\pi \hbar^2 \omega} \textrm{Re}\bigg\{ - \Big( \frac{L}{v} \Big)\int dz  \, \Big[\int  d\vec{R} \, \vec{J}_{fi}^*(\vec{R}, z) e^{i\omega z/v} \Big] e^{-i\omega z/v}\, \cdot \vec{E}_{fi}(\vec{R}_0, z, \omega) \bigg\} \\ 
        & = \frac{1}{\pi \hbar^2 \omega} \textrm{Re}\bigg\{ - \Big( \frac{L}{v} \Big)\int dz  \, \boldsymbol{\mathcal{J}}^{*}_{fi} \, e^{-i\omega z/v}\, \cdot \vec{E}_{fi}(\vec{R}_0, z, \omega) \bigg\} \\
        & = \frac{1}{\pi \hbar^2 \omega} \textrm{Re}\bigg\{ - \Big( \frac{L}{v} \Big) |\boldsymbol{\mathcal{J}}_{fi}| \int dz  \, \boldsymbol{\hat{\mathcal{J}}}^{*}_{fi} \, e^{-i\omega z/v}\, \cdot \vec{E}_{fi}(\vec{R}_0, z, \omega) \bigg\},
    \end{aligned}
    \label{Gamma_loss_fi_nbw}
\end{equation}
where $\boldsymbol{\mathcal{J}}_{fi} = \int d\mathbf{R} \, \boldsymbol{\mathbf{J}}_{fi}(\mathbf{R}) \,  e^{-i \omega z/v}= ({i \hbar e}/{2m})\{ \big[ \braket{\Psi_f | \nabla_{\perp}|\Psi_i} - \braket{\Psi_i | \nabla_{\perp}|\Psi_f}^* \big] + i (2 k_i - q_\parallel)  \braket{\Psi_f | \Psi_i} \hat{\vec{z}}\}$.

\subsection{Recovery of Quasistatic Expressions}

This section outlines the recovery of the energy-resolved loss probability given in Ref. \cite{lourencco2021optical}, where the electron-target interaction is treated quasistatically. In the $c \to \infty$ limit, the Green dyadic is related to the screened interaction $W(\mathbf{x}, \mathbf{x}', \omega)$ by $4 \pi \omega^2 \tensor{\mathbf{G}}(\mathbf{x}, \mathbf{x}', \omega) = \nabla \nabla' W(\mathbf{x}, \mathbf{x}', \omega)$ \cite{lourencco2021optical}. Beginning from Eq. \eqref{gamma_fi_form1} and invoking the non-recoil approximation as well as Eq. \eqref{G_property}, the loss function can be written as
\begin{equation}
    \begin{aligned}
        \Gamma_{fi}(\omega) &= -\frac{4L^2}{\hbar^2 v^2} \textrm{Im}\bigg\{ \int d\vec{x} \, d\vec{x}' \, \vec{J}_{fi}^*(\vec{x}) \cdot \tensor{\mathbf{G}}(\vec{x}, \vec{x}', \omega) \cdot\vec{J}_{fi}(\vec{x}') \bigg\}, \\
        &= -\frac{4 L^2}{\hbar^2 v^2} \textrm{Im}\bigg\{ \int d\vec{x} \, d\vec{x}' \, \vec{J}_{fi}^*(\vec{x}) \cdot \bigg[ \frac{1}{4 \pi \omega^2} \nabla \nabla' W(\mathbf{x}, \mathbf{x}', \omega)  \bigg] \cdot\vec{J}_{fi}(\vec{x}') \bigg\}, \\
    \end{aligned}
\end{equation}
which can be integrated twice by parts to obtain 
\begin{equation}
    \begin{aligned}
        \Gamma_{fi}(\omega) = -\frac{4 L^2}{\hbar^2 v^2} \frac{1}{4 \pi \omega^2} \textrm{Im}\bigg\{ \int d\vec{x} \, d\vec{x}' \, \big[ \nabla \cdot \vec{J}_{fi}^*(\vec{x}, \omega) \big] W(\mathbf{x}, \mathbf{x}', \omega) \big[\nabla' \cdot  \vec{J}_{fi}(\vec{x}', \omega) \big] \bigg\}. \\
    \end{aligned}
\end{equation}
In the case of harmonic time dependence, the continuity equation $\dot{\rho}_{fi}(\mathbf{x}, t) = - \nabla \cdot \vec{J}_{fi}(\mathbf{x}, t) $ becomes $-i \omega_{if}\rho_{fi}(\mathbf{x}) = - \nabla \cdot \vec{J}_{fi}(\mathbf{x})$, with $\rho_{fi}(\mathbf{x}) = - e \psi_{f}^{*}(\mathbf{x})\psi_{i}(\mathbf{x}) = (- e/L) \Psi_{f}^{*}(\mathbf{x})\Psi_{i}(\mathbf{x})e^{i q_{\parallel} z}$. This allows the loss function to be rewritten as
\begin{equation}
    \begin{aligned}
        \Gamma_{fi}(\omega) = \frac{e^2}{\pi \hbar^2 v^2} \int d\vec{x} \, d\vec{x}' \, \Psi_{f}(\mathbf{x})\Psi_{i}^{*}(\mathbf{x}) e^{-i q_{\parallel} z} \textrm{Im}\bigg\{ - W(\mathbf{x}, \mathbf{x}', \omega) \bigg\} \Psi_{f}^{*}(\mathbf{x}')\Psi_{i}(\mathbf{x}')e^{i q_{\parallel} z'}, \\
    \end{aligned}
\end{equation}
which is equivalent to Eq. 1 of Ref. \cite{lourencco2021optical}.


\subsection{Equivalence between $\hat{\boldsymbol{\mathcal{J}}}^{\perp}_{fi} $ and $ \hat{\bf d}^{\perp}_{fi}$}

Ref. \cite{lourencco2021optical} identifies $\hat{\mathbf{d}}^{\perp}_{fi}$ as an OPA in inelastic free electron scattering processes, where $\mathbf{d}^{\perp}_{fi} = -\braket{\Psi_f | e\mathbf{x}_{\perp}|\Psi_i}$. Within the small $w_0$ narrow beam limit,  
\begin{equation}
    \begin{aligned}
        \boldsymbol{\mathcal{J}}^{\perp}_{fi} &= \int d\mathbf{R}\boldsymbol{\mathbf{J}}^{\perp}_{fi}(\mathbf{R}) e^{-i \omega z/v} \\
        &= ({i \hbar e}/{2m}) \big[ \braket{\Psi_f | \nabla_{\perp}|\Psi_i} - \braket{\Psi_i | \nabla_{\perp}|\Psi_f}^* \big]  \\
        &= ({i \hbar e}/{m})  \braket{\Psi_f | \nabla_{\perp}|\Psi_i}.  \\
    \end{aligned}
\end{equation}
Suppose the initial transverse state is the Gaussian state $\Psi^{\textrm{HG}}_{00}(\mathbf{x}_{\perp})$ and the final state resides on the surface of the Poincar{\'e} sphere shown Figure 2f in the main text. Using the fact that $\nabla_{\perp} \Psi^{\textrm{HG}}_{00}(\mathbf{x}_{\perp}) = (-2/w_0^2) \mathbf{x}_{\perp} \Psi^{\textrm{HG}}_{00}$, one finds $\hat{\boldsymbol{\mathcal{J}}}^{\perp}_{fi} \parallel \hat{\mathbf{d}}^{\perp}_{fi}$.

\section{Numerical Calculations}\label{section_simulations}
\begin{figure}
    \centering
    \includegraphics[scale=1.0]{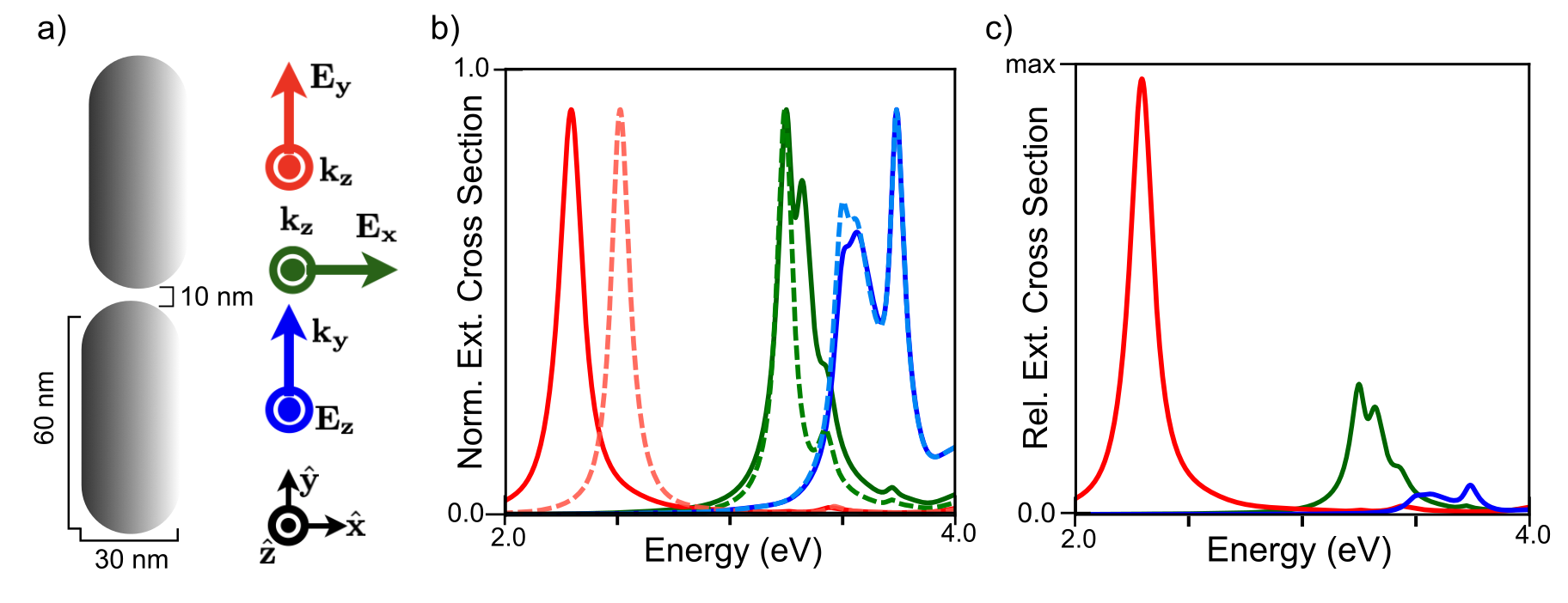}
    \caption{Schematic and optical extinction spectra of Ag rod dimer system considered in Figure 3 of the main text. a) Schematic of the 30 nm $\times$ 60 nm $\times$ 15 nm rod dimer with color-coded diagrams indicating the excitation conditions for panels b) and c). b) Normalized optical extinction cross section spectra for the rod dimer (solid) and single rod (dashed) systems. c) Relative optical extinction cross section spectra for the rod dimer in (b).}
    \label{SI_rod_optical}
\end{figure}

Figure 3 in the main text presents numerically calculated phase-shaped EEL observables for an Ag rod dimer system comprised of two 30 nm $\times$ 60 nm $\times$ 15 nm rods with a 10 nm surface-to-surface gap along the $\hat{\mathbf{y}}$ axis (Figure S\ref{SI_rod_optical}a). Figure S\ref{SI_rod_optical}b shows normalized optical extinction cross section spectra $\sigma_{\textrm{ext}}(\omega)$ calculated using DDSCAT \cite{draine2008discrete} for the rod dimer (solid) and individual rod (dashed) systems excited by plane wave light polarized along $\hat{\mathbf{x}}$ (green), $\hat{\mathbf{y}}$ (red), and $\hat{\mathbf{z}}$ (blue). The amplitudes of the dimer cross section spectra in panel (b) relative to the largest (red) are shown in Figure S\ref{SI_rod_optical}c.

As described in the main text and derived in Section {\ref{section_Derivation_MTEq1}}, both ${\partial^2 \sigma}/{\partial E_{if} \partial \Omega }$ and $\Gamma_{fi}$ observables are dictated by the state- and frequency-resolved loss rate
\begin{equation}
    \begin{aligned}
        w^{\textrm{loss}}_{fi} (\omega ) = \frac{-8 \pi}{\hbar} \textrm{Im}\bigg\{ \int d\vec{x} \, d\vec{x}' \, \vec{J}_{fi}^*(\vec{x}) \cdot \tensor{\mathbf{G}}(\vec{x}, \vec{x}', \omega) \cdot\vec{J}_{fi}(\vec{x}') \bigg\}   \delta (\omega - \omega_{if}). \\
    \end{aligned}
\end{equation}

\begin{figure}
    \centering
    \includegraphics[scale=1.0]{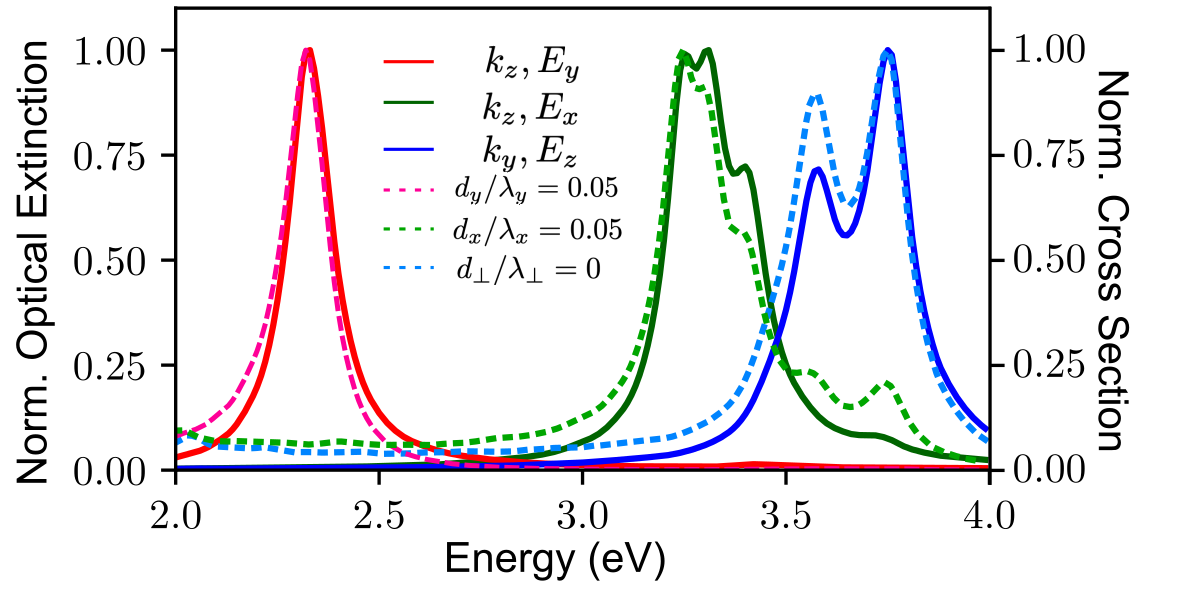}
    \caption{Optical extinction and inelastic electron scattering double differential cross section spectra. Solid lines are normalized optical extinction spectra reproduced from Figure S\ref{SI_rod_optical}. Dashed traces are normalized $\partial^2 \sigma /\partial E_{if} \partial \Omega$  extracted from main text Figure 3a,b within the dipole limit at $d/\lambda_\perp = 0.05$.}
    \label{SI_rod_pwcomp}
\end{figure}

Using the coupled-dipole propagator identity in Eq. \eqref{G_exp_identity}, the portion inside the imaginary operation can be expressed as 
\begin{equation}
    \begin{aligned}
        \int d{\bf x} \int d{\bf x}'~ & \vec{J}_{fi}^{*}(\vec{x})  \cdot \tensor{\bf G}(\vec{x}, \vec{x}', \omega) \cdot \vec{J}_{fi}(\vec{x}') = \\
        &= \int d{\bf x} \int d{\bf x}'~ \vec{J}_{fi}^{*}(\vec{x})  \cdot \sum_{jj'}\tensor{\bf G}_0(\vec{x}-\vec{x}_j, \omega) \bigg( \tensor{\boldsymbol{\alpha}}^{-1} - \tensor{\bf G}_{0} \bigg)_{jj'}^{-1} \tensor{\bf G}_0(\vec{x}_j'-\vec{x}', \omega) \cdot \vec{J}_{fi}(\vec{x}') \\
        & = \int d{\bf x} ~ \vec{J}_{fi}^{*}(\vec{x})  \cdot \sum_{j}\tensor{\bf G}_0(\vec{x}-\vec{x}_j, \omega) \cdot \sum_{j'}\bigg( \tensor{\boldsymbol{\alpha}}^{-1} - \tensor{\bf G}_{0} \bigg)_{jj'}^{-1} \cdot \int d{\bf x}' \tensor{\bf G}_0(\vec{x}_j'-\vec{x}', \omega) \cdot \vec{J}_{fi}(\vec{x}') \\
        & =  \sum_j \frac{v}{L} \bigg(\frac{1}{4 \pi i \omega} \bigg) \vec{E}^{0*}_{fi}(\vec{x}_j, \omega) \cdot \sum_{j'}\bigg( \tensor{\boldsymbol{\alpha}}^{-1} - \tensor{\bf G}_{0} \bigg)_{jj'}^{-1} \cdot \frac{v}{L}\bigg(\frac{1}{-4 \pi i \omega} \bigg) \vec{E}^{0}_{fi}(\vec{x}_{j'}, \omega) \\
        & = \bigg(\frac{v}{L}\bigg)^2\bigg( \frac{1}{4 \pi \omega} \bigg)^2 \sum_j \vec{E}^{0*}_{fi}(\vec{x}_j) \cdot \vec{p}^{j}_{fi}, \\
    \end{aligned}
    \label{wfi_pind}
\end{equation}
where the the induced dipole moments $\vec{p}^{j}_{fi}$ are 
\begin{equation}
    \vec{p}^{j}_{fi} = \sum_{j'}\bigg( \tensor{\boldsymbol{\alpha}}^{-1} - \tensor{\bf G}_{0} \bigg)_{jj'}^{-1} \cdot \vec{E}^{0}_{fi}(\vec{x}_{j'}).
\end{equation}
We modified the $e$-DDA discrete dipole code \cite{Bigelow2012-pq, bigelow2013signatures} to accommodate the transition fields $\vec{E}^{0}_{fi}$ described in Sections \ref{ssection_OPAs_LM} and \ref{ssection_OPAs_OAM} relevant to the transverse state transitions depicted in Figure 2 of the main text. Once the induced dipole moments are determined using $e$-DDA, the $\partial^2 \sigma /\partial E_{if} \partial \Omega$ and $\Gamma_{fi}$ observables are evaluated using Eq. \eqref{wfi_pind}. Figure S\ref{SI_rod_pwcomp} shows $\partial^2 \sigma /\partial E_{if} \partial \Omega$ for the rod dimer system considered in Figure S\ref{SI_rod_optical} for $\mathbf{q}_{\perp} = {\bf 0}$ (dashed blue) as well as for $\mathbf{q}_{\perp} \ne {\bf 0}$ and $\hat{\mathbf{q}}$ oriented along $\hat{\mathbf{x}}$ (red dashed) and $\hat{\mathbf{y}}$ (green dashed) at observation angles $\theta$ (see main text Figure 3) corresponding to $d/\lambda_{\perp} = 0.05$. The normalized optical extinction cross section spectra $\sigma_{\textrm{ext}}(\omega)$ from Figure S\ref{SI_rod_optical}b (solid) are included for comparison.

\newpage
\bibliography{refs.bib}